\def\p{{\bm p}}
\def\q{{\bm q}}
\def\k{{\bm k}}
\def\x{{\bm x}}
\def\y{{\bm y}}
\def\j{{\bm j}}
\def\A{{\bm A}}
\def\Q{{\bm Q}}
\def\tr{\operatorname{tr}}

\def\Re{\operatorname{Re}}
\def\Im{\operatorname{Im}}
\def\grad{{\bm\nabla}}

\def\Res{\operatorname{Res}}

\def\gSG{g_{\rm SG}}
\def\pol{\bar\varepsilon}
\def\kbig{\bar k}
\def\wbig{E}
\def\Aamp{{\cal N}_A}
\def\Acl{A_{\rm cl}}

\def\calGR{{\cal G}_{\rm R}}
\def\calGA{{\cal G}_{\rm A}}

\def\calGRup{{\cal G}^{\rm R}}
\def\calGAup{{\cal G}^{\rm A}}

\def\responseo{\left\langle j^{(3)0}(x) \right\rangle_{\Acl}}

\def\Field{{\cal A}}

\def\xz{x^{\mathsf{3}}}

\def\cth{\operatorname{cth}}
\def\Grrbulk{G_{\rm rr}^{(5)}}
\def\Grrperpbulk{G_{\rm rr\perp}^{(5)}}
\def\GRbulk{G_{\rm R}^{(5)}}
\def\GAbulk{G_{\rm A}^{(5)}}
\def\Gbulk{G^{(5)}}
\def\Sig{\hat\Sigma_\Theta}

\documentclass[prd,preprint,eqsecnum,nofootinbib,amsmath,amssymb,
               tightenlines,dvips]{revtex4}
\usepackage{graphicx}
\usepackage{bm}
\usepackage{amsfonts}

\def\Hl{\operatorname{Hl}}

\begin {document}



\title
    {
      4-point correlators in finite-temperature AdS/CFT:
      jet quenching correlations
    }

\author{
  Peter Arnold and Diana Vaman
}
\affiliation
    {%
    Department of Physics,
    University of Virginia, Box 400714,
    Charlottesville, Virginia 22904, USA
    }%

\date {\today}

\begin {abstract}%
{%
   There has been recent progress on computing real-time
   equilibrium 3-point functions in finite-temperature
   strongly-coupled ${\cal N}{=}4$ super Yang-Mills (SYM).
   In this paper, we show an example of
   how to carry out a similar analysis for a 4-point function.
   We look at the stopping of high-energy ``jets'' in
   such strongly-coupled plasmas and study
   the question of whether, on an event-by-event
   basis, each jet deposits its net charge over a narrow ($\sim 1/T$) or
   wide ($\gg 1/T$) spatial region.
   We relate this question to the
   calculation of a 4-point equilibrium correlator.
}%
\end {abstract}

\maketitle
\thispagestyle {empty}


\section {Introduction}
\label{sec:intro}

Real-time, equilibrium, retarded Green functions are important
to the study of relaxation in finite-temperature field theory.
Through the fluctuation-dissipation theorem, they are directly
related to the dynamics of how small
deviations from equilibrium relax.
In the context of strongly-coupled gauge-theory plasmas with gauge-gravity
duality, 2-point retarded Green functions have been used to
extract viscosity and other hydrodynamic transport coefficients
\cite{SonStarinets,BRSSS}.  3-point retarded Green functions
have been used to
(i) extract the stopping distance for certain types of high-momentum
``jets'' created in the plasma \cite{adsjet,adsjet2,qm11},
and (ii) reproduce earlier results for second-order
hydrodynamic coefficients \cite{AVWXhydro2,SShydro2} and find
the first correction to the large-coupling limits
for such coefficients \cite{SShydro2}.
Finding useful analytic expressions for generic finite-temperature
Green functions can
be a great challenge, but problems of interest are often limiting
cases where the Green functions simplify.  For instance, hydrodynamic
transport coefficients only depend on the low-momentum limit of the
2-point (or 3-point) function, whereas
the 3-point function relevant for the jet
problem is one where two of the three momenta are very high and the other
is low \cite{adsjet}.  In this paper, we will push the analysis of
finite-temperature Green functions in such theories
one point higher: we will discuss a problem, related to jet stopping,
that requires the analysis
of a 4-point function.  In this application, the four momenta involved
will all be either high or low compared to the temperature, which will
make the analysis tractable.

\begin {figure}
\begin {center}
  \includegraphics[scale=0.4]{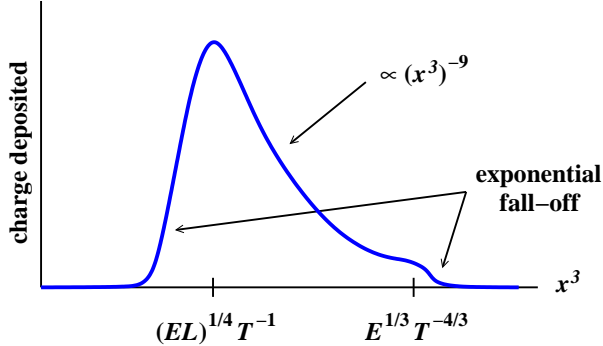}
  \caption{
     \label{fig:charge}
     A qualitative picture of
     the average deposition of charge as a function of $\xz$ for
     jets created by the source described in Sec.\ \ref{sec:source}
     and in ref.\ \cite{adsjet}.
  }
\end {center}
\end {figure}

Here is the problem we study.  We will loosely refer to spatially localized,
high-momentum excitations of the plasma as ``jets.''  Formally, one way
to create such jets is to briefly turn on localized, high-momentum
source terms in the field theory.  In our earlier work \cite{adsjet},%
\footnote{
  See ref.\ \cite{qm11}
  for a very short, breezy, and detail-free overview of ref.\ \cite{adsjet},
}
we
tracked the progress of the resulting jet by measuring the subsequent
evolution $\langle j^0(x) \rangle$ of the density $j^0$ of a conserved
charge carried by the jet.  We found that the jet's charge was deposited
along the jet's trajectory according to a distribution depicted in fig.\
\ref{fig:charge}.
The charge subsequently diffuses hydrodynamically.
In fig.\ \ref{fig:charge},
$L$ is the space-time size of the localized source that we used
to create the jet.  Computing $\langle j^0(x) \rangle$, however, only
gives us the {\it averaged}\/ behavior of a jet.  From this result
alone, one does not know what happens on an event-by-event basis.  Does
each jet (i) deposit over a wide range of distances?  Or, in contrast, does
each jet (ii) stop and deposit its conserved charges in a small localized
region, with fig.\ \ref{fig:charge} representing a probability
distribution for how far the jet travels?
Later work \cite{adsjet2}, which provides a simple way to reproduce
fig.\ \ref{fig:charge} in terms of the trajectories of
classical 5-dimensional particles in AdS$_5$-Schwarzschild,
supports the second interpretation.  It should be possible to
verify this interpretation by examining correlations
$\langle j^0(x) \, j^0(y) \rangle$ of the charge
density in the presence of the source.
If charge deposition is indeed localized for each individual event,
then there should be
no significant correlation between charge deposition at
significantly different distances.

\begin {figure}
\begin {center}
  \includegraphics[scale=0.3]{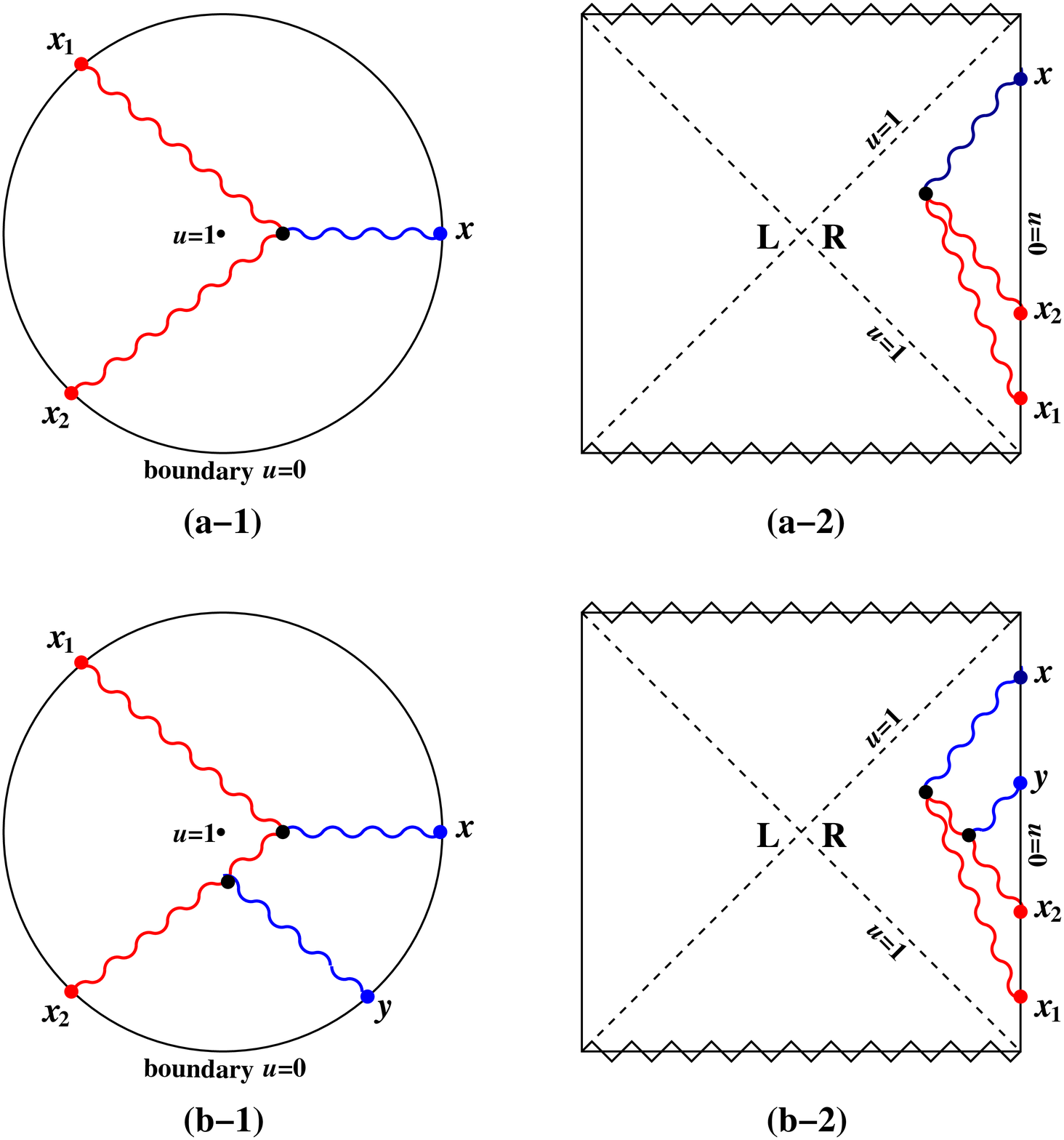}
  \caption{
     \label{fig:witten}
     Examples of Witten diagrams in AdS$_5$-Schwarzschild space
     for (a) a 3-point correlation
     contributing to $\langle j^0(x) \rangle$ in the presence of
     source operators at $x_1$ and $x_2$, and (b) a 4-point correlation
     similarly contributing to $\langle j^0(x) \, j^0(y) \rangle$.
     In both cases, the first diagram depicts traditional Witten
     diagrams drawn in imaginary time, with the boundary ($u{=}0$)
     represented
     as a circle.  The second diagram depicts
     the same diagram in real time superposed atop a Penrose diagram
     of AdS$_5$-Schwarzschild.  For the type of correlators that
     we will examine in this paper, the bulk vertices in the latter case
     are confined by causality to the right-hand quadrant.
     Red lines indicate propagators that have large 4-momenta in our
     calculation (coming from the source) if one transforms to
     4-momentum space, and blue lines (associated with measuring
     hydrodynamic response of charge densities) have low 4-momenta.
     We have used artistic license when drawing the boundary of the
     Penrose diagram with all four sides straight
     \cite{SquarePenrose}.
  }
\end {center}
\end {figure}

As we shall review, the
evolution $\langle j^0(x) \rangle$ of the charge density
in the presence of the source is related to an equilibrium, retarded
3-point function between (i) the measured charge density and
(ii) source operators associated with the creation of the jet.
The corresponding Witten diagram in the gravity dual is depicted
in fig.\ \ref{fig:witten}a.  Correlations such as
$\langle j^0(x) \, j^0(y) \rangle$ will correspond to
equilibrium 4-point functions, and fig.\ \ref{fig:witten}b
gives an example of a corresponding Witten diagram.%
\footnote{
   For a recent discussion of 4-point current correlators at
   {\it zero}\/ temperature, see ref.\ \cite{HattaUeda}.
}
Our goal in this paper is to analyze such diagrams.
We will see later that, for the question of interest here,
the internal line of fig.\ \ref{fig:witten}b is a
bulk-to-bulk propagator with both bulk points located near
the horizon.
There is a relatively simple formula for
the near-horizon bulk-to-bulk propagator, which will be a key
feature in making the analysis tractable.

At the end of
our analysis, we will see that there is indeed no significant
correlation between charge deposition at positions separated
by $\gg 1/T$.

In the next section, we discuss exactly what field theory correlator
we need to measure in order to look for correlations between
jet charge deposition at different locations.
In particular, we will discuss how to define a correlator
$\langle \hat\Theta(x) \, \hat\Theta(y) \rangle_{\rm jet}$ of
``charge deposition'' $\hat\Theta(x)$ and what refinements are
necessary to only capture the physics that we want.
Section \ref{sec:4point}
is devoted to a discussion of exactly what type of
equilibrium 4-point correlator we need in order to evaluate
$\langle \hat\Theta(x) \, \hat\Theta(y) \rangle_{\rm jet}$ and
its relatives and the corresponding Witten diagrams.  We start
with a review of (r,a) notation for Schwinger-Keldysh formalism.
Then we show that we need a 4-point correlator known as
$G_{\rm aarr}$, and we argue that there is only a single Witten
diagram (fig.\ \ref{fig:witten}b with certain specifications
for the causality properties of the propagators)
relevant to the jet deposition question posed in this
introduction.
In preparation for the calculation of that diagram,
section \ref{sec:Grrbulk} examines the equilibrium bulk-to-bulk
correlator $G_{\rm rr}$ in the gravity dual, which will be needed for
the internal line in fig.\ \ref{fig:witten}b.
We show how it can be related to standard bulk-to-boundary propagators.
It turns out that only the case where both bulk points are very close
to the horizon will be relevant, and we
show that the propagators all have a simple structure
in that limit.
With all these preparations in hand, in section \ref{sec:calculate}
we finally calculate the diagram and show that charge deposition
is localized on an event-by-event basis.  Various matters are left
for appendices, and in particular Appendix \ref{app:CCT} relates
our discussion of bulk-to-bulk $G_{\rm rr}$ to a recent discussion by
Caron-Huot, Chesler, and Teaney \cite{CCT}.


\section {General Setup}
\label {sec:general}

\subsection {Notation}

We will use capital roman letters to denote 5-dimensional
space-time coordinates (e.g.\ $I,J=0,1,2,3,5$) and lower-case Greek
letters for 4-dimensional space-time coordinates
(e.g. $\mu,\nu = 0,1,2,3$).

Throughout this paper, we generally refer to the fifth dimension of
AdS$_5$ as $u$, where the boundary is at $u=0$ and the bulk has
$u>0$.  Much of the discussion in the paper will be general but,
when we get down to some specifics later on, we will choose $u$
with AdS$_5$-Schwarzschild metric
\begin {equation}
  ds^2
        =  \frac{R^2}{4} \left[ \frac{(2\pi T)^2}{u}(-f \, dt^2 + d\x^2)
         + \frac{1}{u^2 f} \, du^2 \right] ,
\label {eq:metric}
\end {equation}
where $R$ is the radius of AdS space, $T$ is the temperature,
\begin {equation}
   f \equiv 1 - u^2 ,
\end {equation}
and the horizon is at $u=1$.
At that time, we will also work in units where $2\pi T = 1$.

For 4-vectors $V^\mu$, we define light-cone coordinates
by
\begin {equation}
   V^\pm \equiv V^3 \pm V^0 ,
   \qquad
   V_\pm \equiv \tfrac12 V^\mp = \tfrac12 (V^3 \mp V^0) .
\end {equation}

When writing integrals over 4-momenta, we will use the shorthand
notation
\begin {equation}
   \int_Q \cdots \equiv \int \frac{d^4Q}{(2\pi)^4} \cdots .
\end {equation}


\subsection {The Source}
\label {sec:source}

One could study this problem by creating a ``jet'' with most any
localized high-energy source and then tracking
any conserved charge density.  The calculation is easier with
some choices than others, and here we will follow
ref.\ \cite{adsjet} and study the propagation of R charge
density, using a source that creates R charge.

To later motivate our calculation of jet charge-charge correlations,
it will be convenient to discuss sources that create jets with
R charge $\gg 1$.  On the other hand, it is also convenient to keep
the detailed structure of the calculations here as close as possible
to our previous calculations in ref.\ \cite{adsjet}, where the
source created jets carrying a single unit of R charge.
As a compromise, in the main text we will use the single-charge
source of ref.\ \cite{adsjet}, described below, when writing down
specific formulas involving the source.
But little depends on this, and in Appendix \ref{app:Rcharge},
we discuss how the formulas generalize to cases
where the R charge of the jet is large compared to 1.

So, following ref.\ \cite{adsjet}, we modify the
4-dimensional field theory
Lagrangian by
\begin {subequations}
\label {eq:SOURCE}
\begin {equation}
   {\cal L} \to {\cal L} + j_\mu^a \Acl^{a\mu} ,
\label {eq:Lsource}
\end {equation}
where $j_\mu^a$ are the SU(4) R-charge currents of the
theory and $\Acl$ is a classical external source.
We choose the external source to have the form of (i) a high-energy
plane wave $e^{i\kbig\cdot x}$ times (ii) a smooth, slowly varying,
real-valued envelope
function $\Lambda_L(x)$ localizing the source to a space-time
region of size $L$.  Exactly as in ref.\ \cite{adsjet}, we take
\begin {equation}
   \Acl^{\mu}(x)
   = \pol^\mu \Aamp \Bigl[
       \frac{\tau^+}{2} \, e^{i \kbig\cdot x} +
       {\rm h.c.}
      \Bigr] \, \Lambda_L(x) ,
\label{eq:source}
\end {equation}
\end {subequations}
where
\begin {equation}
   \kbig^\mu = (\wbig,0,0,\wbig)
\label {eq:kbig}
\end {equation}
is a very large light-like 4-momentum with frequency
$\wbig \gg T$;
$\Aamp$ is an arbitrarily small source amplitude;
$\pol$ is a transverse linear polarization, such as
\begin {equation}
   \pol^\mu = (0,1,0,0) ;
\end {equation}
and $\tau^i$ are Pauli matrices
for any SU(2) subgroup of the
SU(4) R-symmetry, with
\begin {equation}
   \tau^\pm = \tau^1 \pm i \tau^2 .
\label {eq:taupm}
\end {equation}
A simple example of an appropriate envelope function would be
\begin {equation}
   \Lambda_L(x)
   = e^{-\frac12 (x_0/L)^2} e^{-\frac12 (x_3/L)^2} .
\label {eq:Genvelope}
\end {equation}
$L$ should be chosen large compared to $1/\wbig$, so that the
momentum components in the source are all close to
(\ref{eq:kbig}), but small compared to the large stopping distance
that we wish to study.
The creation of R charge by the source (\ref{eq:SOURCE}) is
analogous to a decay $W^+ \to u \bar d$ of a very high momentum
$W$ boson creating a localized excitation with isospin in
a standard-model quark-gluon plasma.


\subsection {The Measurement: Previous Work}
\label {sec:previous}

In previous work \cite{adsjet}, we measured the late-time behavior of
\begin {equation}
   \langle j^{(3)0}(x) \rangle_{A_{\rm cl}}
\label {eq:j30}
\end {equation}
for a system that started in thermal equilibrium.
The superscript ``(3)'' indicates the R current associated with
$\tau^3/2$ in the SU(2) subgroup referenced by (\ref{eq:taupm}).
The subscript ``$A_{\rm cl}$'' indicates that the expectation is
taken with the source term (\ref{eq:Lsource}) present in the
Lagrangian.  As reviewed in ref.\ \cite{adsjet}, expanding to leading
order in the small-amplitude source gives
\begin {equation}
  \langle j^{(3)\mu}(x) \rangle_{A_{\rm cl}}
  =
  \tfrac12
  \int d^4x_1 \> d^4x_2 \>
  G_{\rm aar}^{(ab3)\alpha\beta\mu}(x_1,x_2;x) \,
  A^a_{\alpha,\rm cl}(x_1) \,
  A^b_{\beta,\rm cl}(x_2) ,
\label {eq:jss}
\end {equation}
or equivalently
\begin {equation}
  \langle j^{(3)\mu}(x) \rangle_{A_{\rm cl}}
  =
  \tfrac12
  \int_{Q_1Q_2Q}
  G_{\rm aar}^{(ab3)\alpha\beta\mu}(Q_1,Q_2;Q) \,
  A^{a*}_{\alpha,\rm cl}(Q_1) \,
  A^{b*}_{\beta,\rm cl}(Q_2) \,
  e^{iQ\cdot x}
  (2\pi)^4 \delta^{(4)}(Q_1+Q_2+Q) ,
\label {eq:jssQ}
\end {equation}
where $G_{\rm aar}$ is the retarded, equilibrium 3-point Green function
associated with source operators $j\cdot A_{\rm cl}$ at $x_1$ and
$x_2$ and the measurement operator $j^{(3)\mu}$ at $x$.

In this paper, we will use (r,a) notation to indicate different
orderings of operators in $n$-point thermal Green functions.
We review and summarize this notation (and fix our
normalization conventions) in section \ref{sec:rarules},
but for the moment it's enough to
remark that $G_{\rm aa\cdots ar}(x_1,x_2,\cdots,x_{n-1};x)$
denotes an $n$-point retarded Green function where $x$ is the
measurement point and $x_1,\cdots, x_{n-1}$ are source points.

After a high energy jet stops and thermalizes in the plasma, its
charge density diffuses out from the place
where the jet stopped.
What we most directly want to know is where the
charge density was deposited \cite{CJK,adsjet}.
If locally thermalized
charge is deposited with density $\hat\Theta(x)$ in the plasma,
the late time evolution of the charge will be given by the
diffusion equation
\begin {equation}
   (\partial_t - D \grad^2) \, j^0(x) = \hat\Theta(x) ,
\label {eq:ThetaMotivate}
\end {equation}
where \cite{Rdiffusion}
\begin {equation}
   D = \frac{1}{2\pi T}
\label {eq:Dvalue}
\end {equation}
is the R-charge diffusion constant.  We can turn (\ref{eq:ThetaMotivate})
around into an operator definition of the charge deposition:%
\footnote{
   The $\bar {\cal Q} \, \Theta(x)$ of ref.\ \cite{adsjet} is
   $\langle \hat\Theta(x) \rangle_{A_{\rm cl}}$ in our notation
   here.
}
\begin {equation}
    \hat\Theta(x) \equiv
    (\partial_t - D \grad^2) \,
    j^0(x) .
\label {eq:Theta}
\end {equation}
Here and throughout this paper, we will often denote the
measurement operator $j^{(3)\mu}$ by simply $j^\mu$ when there
is small chance of confusion.

The charge deposition function $\hat\Theta(x)$ is only significant
very close to the line $x^0 = x^3$ of travel of the high-energy
jet.  We shall not concern ourselves with the detailed profile
of the width of this function.  And so, following ref.\ \cite{adsjet},
we will write
\begin {subequations}
\label{eq:SigmaIntro}
\begin {equation}
   \langle \hat\Theta(x) \rangle_{A_{\rm cl}}
   \simeq
   \delta_L(x^-) \,
   \langle \Sig(t{=}\infty,\x) \rangle_{A_{\rm cl}} ,
\end {equation}
where
\begin {equation}
   \Sig(x) \equiv
   \int_{-\infty}^t dt' \> \hat\Theta(t',\x) ,
\label {eq:Sigma}
\end {equation}
\end {subequations}
and the $L$ subscript on $\delta_L(x^-)$ indicates that the
$\delta$-function approximates a function whose width is of order
$L$.

\begin {figure}
\begin {center}
  \includegraphics[scale=0.4]{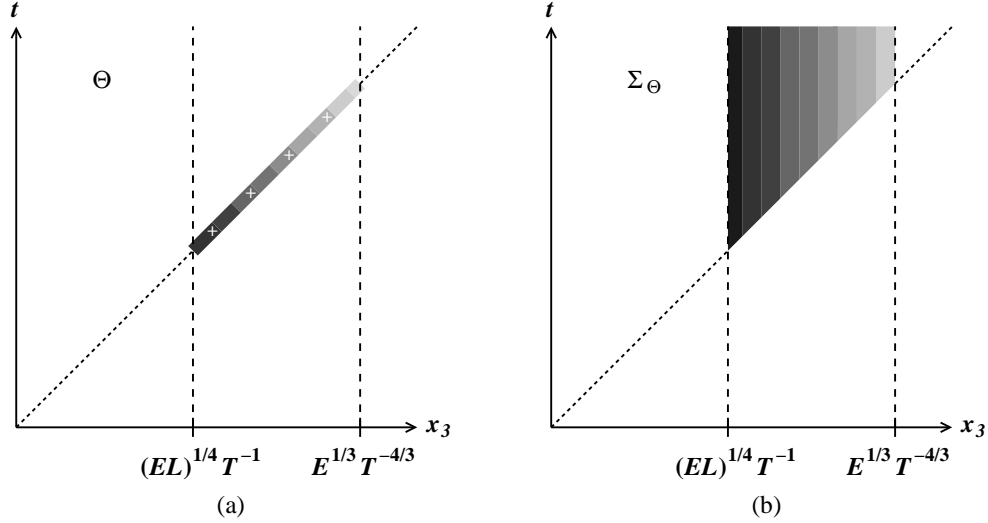}
  \caption{
     \label{fig:Sigma}
     A cartoon of the space-time distribution of
     (a) $\langle \hat\Theta(x) \rangle$ and
     (b) $\langle \hat\Sigma_\Theta(x) \rangle$
     in the region of $x^3$ where there is net charge deposition.
     (See the discussion of fig. 9 of ref.\ \cite{adsjet} for a more detailed
     description.)
  }
\end {center}
\end {figure}

See ref.\ \cite{adsjet} for qualitative discussion of what
the functions $\hat\Theta(x)$ and $\Sig(x)$ look like.
Here we will just summarize that the fact that net charge
deposition is restricted to a region near the light cone,
as shown by the cartoon picture of $\langle\hat\Theta(x)\rangle$ in
Fig.\ \ref{fig:Sigma}a, means that the time-integrated charge
deposition $\langle\hat\Sigma_\Theta(x)\rangle$ will be constant in
time for times well inside the light-cone, as depicted in
Fig.\ \ref{fig:Sigma}b.  We can summarize this constancy
as
\begin {equation}
   \langle \Sig(t{=}x^3{+}\epsilon,x^3) \rangle_{A_{\rm cl}}
   \simeq
   \langle \Sig(t{=}\infty,x^3) \rangle_{A_{\rm cl}}
\label {eq:whereSigma}
\end {equation}
whenever $\epsilon \gg L$.  The important thing to remember is
that even though we will find it convenient to talk about and calculate
$\Sig(t{=}\infty,x^3)$,
that quantity is determined by charge deposition that takes place
at $t \simeq x^3$, not at $t=\infty$.

The great advantage of phrasing the problem in terms of the
$t{\to}\infty$ limit of $\Sig(x)$ is that the
calculation in the gravity dual turns out to
then only depend on the behavior
of the corresponding 5-dimensional excitation at late times, when it has
fallen very close to the event horizon of the black hole.  For the
purpose of calculation, 5-dimensional propagators in Witten diagrams
turn out to be much simpler and easier to deal with when the bulk
points are very close to the horizon.

  It's important to note that most of the time, a localized,
small-amplitude source will not have any effect at all on
the plasma.  It is only on rare occasions, proportional to the
square $|{\cal A}|^2$ of the source amplitude, that the source will
create an excitation.
When it does, the excitation will have the quantum numbers of the
source.  The jets will have energy and momentum $\simeq E$.
Their R charge $e_{\rm jet}$ will be the R charge of the source
operator,\footnote{
  Here and throughout, when we talk about the R charge $e_{\rm jet}$
  of a jet, we mean the charge corresponding to the specific
  U(1) subgroup of SU(4) that we have chosen for our measurement
  operator, i.e.\ the charge corresponding to what we have labeled
  $j^{(3)0}$ in (\ref{eq:j30}) and (\ref{eq:calQ}).
}
which is $e_{\rm jet}{=}1$ for the specific
choice (\ref{eq:source}) and may take other values for the
examples discussed in Appendix \ref{app:Rcharge}.
The average charge created in response to the source,
\begin {equation}
   {\cal Q} \equiv
   \int d^3x \> \responseo
   \biggl|_{x^0 \gg L} ,
\label {eq:calQ}
\end {equation}
is therefore $e_{\rm jet}$ times the probability ${\cal P}_{\rm jet}$
that a jet is created.
In general, we will be interested in average charge densities and
charge deposition {\it in the case that a jet has actually been
created}.  We can factor out the probability of creating the
jet in the first place by defining expectations
\begin {equation}
   \langle \cdots \rangle_{\rm jet}
   \equiv
   \frac{\langle \cdots \rangle_{A_{\rm cl}}}{{\cal P}_{\rm jet}}
   =
   \frac{\langle \cdots \rangle_{A_{\rm cl}}}{{\cal Q}/e_{\rm jet}}
   \,.
\label {eq:anglejet}
\end {equation}
So, in the notation of this paper, the average charge deposition of
our jets is given by
\begin {equation}
   \langle \hat\Theta(x) \rangle_{\rm jet}
   =
   \frac{\langle \hat\Theta(x) \rangle_{A_{\rm cl}}}{{\cal Q}/e_{\rm jet}} \,.
\end {equation}

Finally, at a practical level, the calculation
of the Green function in (\ref{eq:jss}) took advantage of
approximations based on the momentum scales relevant to the problem.
Since we measure where the charge is deposited by its subsequent
diffusion, and since diffusion is a hydrodynamic process that takes
place on distance and time scales large compared to the mean-free
path and so large compared to $1/T$, we do not need to resolve the
structure of $\langle j^{(3)0}(x)\rangle$ on scales as small as
$1/T$.  In particular, we do not need to know the integrand of
(\ref{eq:jssQ}) except for when the components of $Q$ (the 4-momentum
conjugate to the measurement position $x$) are small compared to $T$.
In contrast, the momenta $Q_1$ and $Q_2$ associated with the source
factors in (\ref{eq:jssQ}) are both large compared to $T$ because
we have chosen a high-energy/high-momentum source.  That means
that, after Fourier transforming the boundary points, the (red)
propagators associated with the source points $x_1$ and $x_2$ in the
Witten diagram of fig.\ \ref{fig:witten}a will have high
4-momentum, and so can be treated in a WKB-like approximation,
whereas the (blue) propagator associated with the measurement point
$x$ will have low 4-momentum, which also allows for simplification.
Similar simplifications will occur when calculating 4-point
correlators such as fig.\ \ref{fig:witten}b.


\subsection {The Measurement: This Paper}

\subsubsection {A first attempt}

\begin {figure}
\begin {center}
  \includegraphics[scale=0.4]{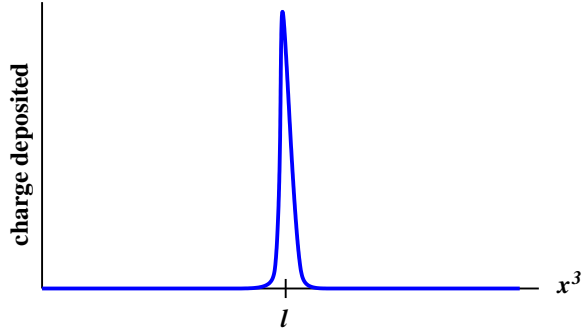}
  \caption{
     \label{fig:narrow}
     A qualitative picture of charge deposition in a single event if
     charge deposition is very localized on an event-by-event basis.
     In this case, the location $l$ of where the charge is deposited
     would have a probability distribution given by fig.\ \ref{fig:charge}.
  }
\end {center}
\end {figure}

To motivate our calculation, imagine that we used a source that
creates jets with large R charge $(e_{\rm jet} \gg 1)$.
Then ask whether, in a single typical event, the spatial
distribution of deposited charge looks like
(i) fig.\ \ref{fig:charge} or (ii) fig.\ \ref{fig:narrow}.
In the latter case, fig.\ \ref{fig:charge} would give the
probability distribution for the position $l$ in fig.\
\ref{fig:narrow}.  Naively, a simple way to distinguish these two
cases would be to measure the average correlation
\begin {equation}
   \bigl\langle \hat\Theta(x) \, \hat\Theta(y) \bigr\rangle_{\rm jet}
   =
   \frac{\bigl\langle \hat\Theta(x) \, \hat\Theta(y)\bigr\rangle_{A_{\rm cl}}}
        {{\cal Q}/e_{\rm jet}}
\end {equation}
for well separated $x$ and $y$.
The idea is that, for well separated $x$ and $y$,
case (i) would give
\begin {equation}
   \bigl\langle \hat\Theta(x) \, \hat\Theta(y) \bigr\rangle_{\rm jet}
   \simeq
   \bigl\langle \hat\Theta(x) \bigr\rangle_{\rm jet} \,
   \bigl\langle \hat\Theta(y) \bigr\rangle_{\rm jet} ,
\label {eq:factorize}
\end {equation}
whereas case (ii) would instead give
\begin {equation}
   \bigl\langle \hat\Theta(x) \, \hat\Theta(y) \bigr\rangle_{\rm jet}
   \simeq
   0
\label {eq:nocorr}
\end {equation}
if $x$ and $y$ were much further apart than the width the peak in
fig.\ \ref{fig:narrow}.

When discussing correlators, there are
operator ordering considerations except in the classical limit.
Note, for instance, that the right-hand side of (\ref{eq:factorize})
is (i) symmetric with respect to $x\leftrightarrow y$ and (ii) real, whereas
the left-hand side is not precisely either.
Factorization such as (\ref{eq:factorize})
would require the imaginary part of the correlator to be small.
The real and imaginary parts of
$\langle \hat\Theta(x) \, \hat\Theta(y) \rangle$
are given respectively by the symmetric combination
$\tfrac12 \langle \{ \hat\Theta(x), \hat\Theta(y) \} \rangle$
and the commutator
$-i \tfrac12 \langle [ \hat\Theta(x), \hat\Theta(y) ] \rangle$.
In this paper, we will focus on the real part, for which the two
cases distinguished above would be
\begin {equation}
   \tfrac12
   \bigl\langle \{ \hat\Theta(x) , \hat\Theta(y) \} \bigr\rangle_{\rm jet}
   \simeq
   \bigl\langle \hat\Theta(x) \bigr\rangle_{\rm jet} \,
   \bigl\langle \hat\Theta(y) \bigr\rangle_{\rm jet}
\label {eq:factorize2}
\end {equation}
and
\begin {equation}
   \tfrac12 \bigl\langle \{ \hat\Theta(x) ,\hat\Theta(y) \}
            \bigr\rangle_{\rm jet}
   \simeq
   0 .
\label {eq:nocorr2}
\end {equation}

As we will see in a moment, the idea for our correlation measurement
will require a little bit of refinement,
but the basic goal of this paper is to demonstrate a refined version of
(\ref{eq:nocorr2}) and so verify that the deposition of charge by a jet
looks like fig.\ \ref{fig:narrow} on an event-by-event basis
with a width $\Delta x^3 \lesssim 1/T$.
We will refer to this as ``localized'' deposition of charge.

Eqs. (\ref{eq:factorize2}) and (\ref{eq:nocorr2}) lay out two logical
extremes for how a correlation might behave, but {\it a priori}\/ the
truth could instead lie somewhere in between.  In particular,
factorization (\ref{eq:factorize2}) is not a very plausible
option for non-local correlations unless $e_{\rm jet}{\gg}1$.
But the simple question of whether the correlation is local or non-local
can be sensibly asked for $e_{\rm jet}{=}1$ as well as $e_{\rm jet}{\gg}1$.
In the main text, for the sake of concreteness and compatibility with
previous work, we will focus on the case $e_{\rm jet}{=}1$
corresponding to the specific source (\ref{eq:source}).
For the sake of completeness, we will discuss the same question for
$e_{\rm jet}{\gg}1$ in Appendix
\ref{app:Rcharge}.
In both cases, we will find locality for the suitably-refined
correlation.

In calculations, we will find it easier to study the correlator
\begin {equation}
     \tfrac12
     \bigl\langle \{ \Sig(x^0{=}\infty,\x),
                     \Sig(y^0{=}\infty,\y) \}
     \bigr\rangle
\end {equation}
of time-integrated charge-deposition (\ref{eq:Sigma}) rather than
$\tfrac 12 \langle \{ \hat\Theta(x), \hat\Theta(y) \} \rangle$
directly.


\subsubsection {Refinement}
\label {sec:refine}

When we later evaluate the correlation
$\tfrac12 \langle \{\hat\Theta(x) , \hat\Theta(y)\} \rangle$
in the gravity dual, we will see that we want to throw out
certain types of diagrammatic contributions which do not
have the causal structure appropriate for the question of how charge
deposition from a jet is correlated.
In this section, we discuss what goes wrong if one includes
all contributions to 
$\tfrac12 \langle \{\hat\Theta(x) , \hat\Theta(y)\} \rangle$,
and we outline a more careful definition of charge deposition
which avoids unwanted contributions and
only includes the physics of interest.

From our definition (\ref{eq:Theta}) of charge deposition
$\hat\Theta(x)$, the correlator we proposed above for study is
\begin {equation}
   \tfrac12
   \bigl\langle \{ \hat\Theta(x) ,\hat\Theta(y) \} \bigr\rangle_{\rm jet}
   =
   (\partial_{x^0} - D \grad_{\x}^2)
   (\partial_{y^0} - D \grad_{\y}^2)
   \tfrac12
   \bigl\langle \{ j^0(x) , j^0(y) \} \bigr\rangle_{\rm jet}
   .
\label {eq:TTdiffuse}
\end {equation}
This would be fine if the only source of charge in the system
contributing to this correlator were the charge deposited by
the jet.  However, even if there were no jet at all,
the equilibrium correlator
$\langle \{ j^0(x), j^0(y) \} \rangle_{\rm eq}$ would be non-vanishing and
would measure the correlation of charge-density fluctuations in
the plasma.  Not only is the equilibrium correlation not
the physics we want to measure for studying jet stopping but, unlike
the case of charge deposition, its time dependence is not
the simple retarded time-dependence of the diffusion equation
implicitly assumed in the introduction (\ref{eq:ThetaMotivate})
of the charge deposition distribution $\hat\Theta(x)$.
To see this, consider that the equilibrium charge density correlation function
$\langle \{ j^0(x), j^0(y) \} \rangle_{\rm eq}$ can be related
to the retarded charge density correlator via the
fluctuation-dissipation theorem:%
\footnote{
  See, for example, the discussion by Kubo \cite{Kubo}.
  Eq.\ (\ref{eq:jjfluctdisp})
  above is a special case of our later (\ref{eq:fluctdisp}).
}
\begin {equation}
   \tfrac12
   \bigl\langle \{ j^0 , j^0 \} \bigr\rangle_{\rm eq}
   =
   -\cth(\tfrac12\beta\omega) \, \Im G_{\rm R}^{00}(\omega,\q) ,
\label {eq:jjfluctdisp}
\end {equation}
where we work in 4-momentum space $(\omega,\q)$.
In the hydrodynamic limit of small frequency and momentum,
$G_{\rm R}^{00} \propto D\q^2/(-i\omega+D\q^2)$ has retarded
diffusive behavior, and (\ref{eq:jjfluctdisp}) gives
\begin {equation}
   \tfrac12
   \bigl\langle \{ j^0 , j^0 \} \bigr\rangle_{\rm eq}
   \simeq
   \frac{2T}{\omega}
   \Im\left( \frac{\chi}{T} \, \frac{D \q^2}{(-i\omega + D \q^2)} \right)
  = \frac{2 \chi D \q^2}{\omega^2 + (D \q^2)^2}
  \label {eq:eqcorr}
  ,
\end {equation}
where $\chi$ is the charge susceptibility.%
\footnote{
  The overall normalization of the right-hand side of
  (\ref{eq:eqcorr}) is determined
  by integrating both sides $d\omega/2\pi$ over all $\omega$ and then taking
  $\q \to 0$.  After this procedure, the left-hand side becomes
  the equal-time expectation
  $\langle (\mbox{total charge})^2 \rangle / \mbox{volume}$, which
  defines the charge susceptibility $\chi$.
  For a more specific formula for the hydrodynamic limit
  of $G_{\rm R}^{00}(\omega,\q)$ for ${\cal N}{=}4$ SYM in particular, see,
  for example, eq.\ (79) in the review article \cite{SonStarinets}.
}
It's useful to also write this as
\begin {equation}
   \tfrac12
   \bigl\langle \{ j^0 , j^0 \} \bigr\rangle_{\rm eq}
   \simeq
   \frac{2\chi}{i\omega} \left[
      \frac{D\q^2}{-i\omega+D\q^2}
      - \frac{D\q^2}{i\omega+D\q^2}
   \right]
   .
\label {eq:corr2}
\end {equation}
There is not only a diffusive pole $-i\omega+D\q^2=0$ but also
a conjugate pole $i\omega+D\q^2=0$ corresponding to its time reversal.
Multiplying both sides by
factors of $-i\omega+D\q^2$ to turn $j^0$'s into $\hat\Theta$'s
would simplify the first term of (\ref{eq:corr2}) but not the second.
The difficulty is that the correlation
represented by
$\bigl\langle \{ j^0 , j^0 \} \bigr\rangle_{\rm eq}$ is
time symmetric, unlike the diffusion operator
$\partial_t - D\grad^2$ and unlike the physics of charge that
is suddenly deposited in the medium by a thermalizing jet and
only subsequently diffuses.
The upshot is that the equilibrium contribution to
$\langle \{ j^0(x), j^0(y) \} \rangle_{\rm eq}$ messes up
our desired physical interpretation of the correlation
(\ref{eq:TTdiffuse}).

There is a simple fix to the problem identified so far: Subtract
out the equilibrium contribution to the charge correlation,
and so study
\begin {equation}
   (\partial_{x^0} - D \grad_{\x}^2)
   (\partial_{y^0} - D \grad_{\y}^2)
   \left[
     \bigl\langle \{ j^0(x) , j^0(y) \} \bigr\rangle_{\rm jet}
   - \bigl\langle \{ j^0(x) , j^0(y) \} \bigr\rangle_{\rm eq}
   \right]
\label {eq:jjsubtract}
\end {equation}
instead of (\ref{eq:TTdiffuse}).

\begin {figure}
\begin {center}
  \includegraphics[scale=0.4]{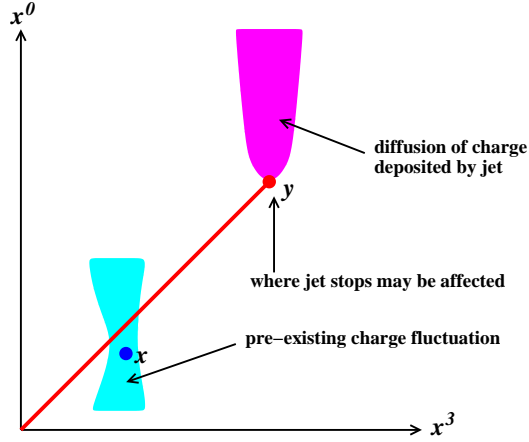}
  \caption{
     \label{fig:bias}
     A cartoon depicting a possible correlation between a
     charge fluctuation in the plasma at $x$ and the deposition
     of the jet's charge at $y$.
  }
\end {center}
\end {figure}

But there is still a problem.  Consider the case where the jet
passes through a region of plasma that contains some equilibrium
fluctuation in the charge density.  It's conceivable that
interaction with that fluctuation, depending on its sign, could
bias the probability that the jet stops earlier or later, as depicted
in fig.\ \ref{fig:bias}.  So we could get a contribution to
(\ref{eq:TTdiffuse}) where the $j^0(x)$, for example, represents
a pre-existing charge fluctuation in the plasma (i.e.\ not
something caused by the jet) and the $j^0(y)$ represents
diffusion from a jet depositing its charge at $y$.
Let's call this a plasma-jet contribution to the correlation.
Such a contribution
would contaminate what we really want to know, which is the correlation
between the jet depositing its charge at $x$ and depositing its charge
at $y$.  Plasma-jet contributions to the correlation would also not
give the retarded time dependence in $x^0$ assumed by the
$(\partial_{x^0} - D \grad_{\x}^2)$ factor in (\ref{eq:TTdiffuse})
for the same physical reason as in the discussion of
the purely equilibrium correlator
$\langle \{j^0(x), j^0(y)\} \rangle_{\rm eq}$.
This vague concern will be made concrete when we later discuss
diagrammatic contributions to (\ref{eq:TTdiffuse}) in the
(r,a) formalism.  We will see that some diagrams correspond to
the physics of the charge deposition correlation that we want, while
other diagrams correspond to plasma-jet correlations that we are
not interested in for the purposes of answering our question about
how charge is deposited.

\begin {figure}
\begin {center}
  \includegraphics[scale=0.4]{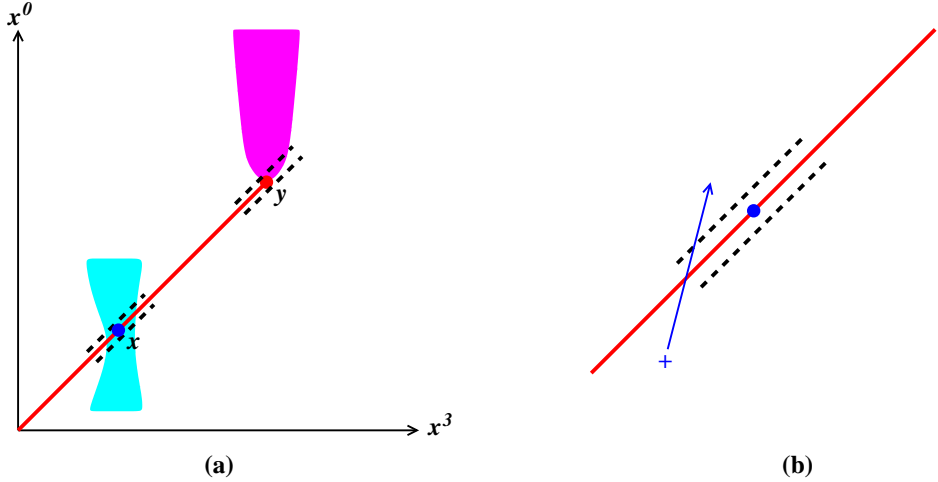}
  \caption{
     \label{fig:planes}
     (a) The hyper-plane segments (dashed lines) used to define
         $\bar\Sigma_\Theta$ corresponding to charge deposition
         near the light-cone points $x$ and $y$.
     (b) An enlarged picture of one set of these plains showing how a
         charge fluctuation from the plasma might sneak between
         a pair of planes through one of the open ends and so
         be wrongly counted in the charge deposition
         measurement.
  }
\end {center}
\end {figure}

Since neither the original correlator
$\tfrac12 \langle \{\hat\Theta(x), \hat\Theta(y)\} \rangle$
nor the subtracted version (\ref{eq:jjsubtract})
exactly capture the physics we want, how can we define something that
does?  The problem is that our definition of charge deposition as given
by (\ref{eq:Theta}) for $\hat\Theta(x)$ is not really a good definition
of charge deposition when we want to study correlators.  We should
define more carefully what it means for the jet to deposit charge
somewhere.  Here is one way to do that.  Until the jet deposits its
charge, we know from previous work that it stays close to the light-cone
$x^0 = x^3$, within a distance $x^- \sim L$ (assuming $L \gtrsim 1/T$).
We wish to define the net charge density deposited by the jet in the
vicinity of some point $\bar x^+$ along the light-cone.
Consider two 3-dimensional hyper-planar strips defined by
\begin {equation}
   x^- = \pm \delta_{\rm plane} ,
   \qquad
   | x^+ - \bar x^+ | \le \ell_{\rm plane} ,
\end {equation}
corresponding to the dashed lines in fig.\ \ref{fig:planes}a.
Choose $\delta_{\rm plane}$ and $\ell_{\rm plane}$ so that
$L \ll \delta_{\rm plane} \ll \ell_{\rm plane} \ll \Delta x^+$,
where $\Delta x^+$ is whatever scale at which one wants to resolve
differences between stopping distances.
$\Delta x^+$ should definitely be
chosen small compared to the typical stopping distance $(EL)^{1/4}/T$.
Consider the space-time region bounded by the two planes and
approximately define the charge deposited at $\bar x^+$ as
the difference between the charge leaving that region through the
top plane and entering it through the bottom plane:
\begin {align}
   \bar\Sigma_\Theta(\infty,x^3{=}\tfrac12 \bar x^+)
   &\equiv
   \frac{1}{\ell_{\rm plane} V_\perp}
   \int dS_\mu \> j^\mu
\nonumber\\
   &=
   \frac{1}{\ell_{\rm plane}}
   \int_{\bar x^+-\frac12 \ell_{\rm plane}
      }^{\bar x^++\frac12 \ell_{\rm plane}}
   \frac{dx^+}{2} \int \frac{d^2\x^\perp}{V_\perp}
   \Bigl[
      j^-\bigl( \frac{x^+-\delta_{\rm plane}}{2}, \x^\perp,
                \frac{x^++\delta_{\rm plane}}{2} \bigr)
\nonumber\\ &\hspace{14em}
      -
      j^-\bigl( \frac{x^++\delta_{\rm plane}}{2}, \x^\perp,
                \frac{x^+-\delta_{\rm plane}}{2} \bigr)
   \Bigr] ,
\label {eq:NewSigma}
\end {align}
where the $dS$
integral in the first line is over both hyper-plane segments, with
$dS_\mu$ pointing outward from the region between,
and $V_\perp$ is the area of the transverse dimensions $\x_\perp$.%
\footnote{
   Remember that we have chosen to set up a transverse
   translationally invariant problem, so $V_\perp$ is just an infinite
   (and uninteresting) normalization factor.  Eq.\ (\ref{eq:NewSigma})
   really defines the $\x_\perp$-averaged time-integrated charge
   deposition function.  We could have equally well defined the
   symbol $\Sigma_\Theta$
   throughout this paper to be the $\x_\perp$-integrated (rather than
   averaged) time-integrated
   charge density, in which case there would be no $1/V_\perp$ factors
   in (\ref{eq:NewSigma}).
}
The jet sneaks into
the region, without crossing either plane, through the lower-left opening
between the planes in fig.\ \ref{fig:planes}a.  If it does not thermalize
there, then it sneaks out again, uncounted, through the upper-right
opening.  If some or all of its charge thermalizes in the region, that
deposition of charge will be counted by (\ref{eq:NewSigma}).
This definition solves the problem of jet-plasma correlations because
a pre-existing charge fluctuation in the vicinity of $x$ will give
a canceling contribution to (\ref{eq:NewSigma}) between passing
through the lower and upper planes, as depicted in fig.\ \ref{fig:planes}a.
There will
be errors in this procedure because of edge effects: some fluctuation in
the equilibrium charge density might sneak in one side and be counted,
as shown in fig.\ \ref{fig:planes}b.  But such effects will scale away
as one makes $\ell_{\rm plane}$ larger and so can be disentangled.

We use the bar over $\Sigma_\Theta$ in (\ref{eq:NewSigma}) to distinguish
our refined version of $\hat\Sigma_\Theta(\infty,x^3)$
from our original definition
(\ref{eq:Sigma}).  The bar can also crudely be thought of indicating
that this definition averages $\hat\Sigma_\Theta(\infty,x^3)$ over a distance
of order $\ell_{\rm plane}$, and so introduces some smearing.
Such smearing is inconsequential as
long as one keeps $\ell_{\rm plane}$ small compared to the
differences in stopping distances that one wishes to resolve.
Finally, we remind the reader that the quantity $\Sigma_\Theta(\infty,x^3)$
arises from charge that is deposited near the light-cone
$x^0 \simeq x^3$ and not at $x^0{=}\infty$.
See (\ref{eq:whereSigma}).

\begin {figure}
\begin {center}
  \includegraphics[scale=0.4]{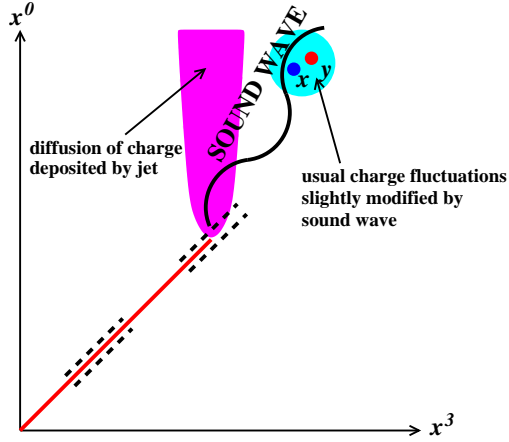}
  \caption{
     \label{fig:sound}
     A jet travels along the light cone and then deposits its charge,
     which diffuses, as well as sound energy, which propagates.
     The local oscillation of temperature associated with the sound
     wave can then modify the local correlations between $j^0(x)$
     and $j^0(y)$ at, for example, the points shown.  Such a change in
     the correlation would contribute to (\ref{eq:jjsubtract}) but
     is invisible to the refined measurement (\ref{eq:NewSigma}), which
     is defined specifically for $x$ and $y$ on the light cone and
     only involves
     $j^0$ at points close to the light cone, such as at the dashed lines.
  }
\end {center}
\end {figure}

Consider now yet another possible source of
$\langle \{ j^0(x), j^0(y) \} \rangle$ correlation.
When the jet stops, it deposits energy and momentum into the
medium, which can then propagate away in the form of sound waves.
These sound waves locally increase and decrease the temperature
and pressure as they pass by.
A small change $\delta T$ in local temperature will give a small, local
change to the contribution to $\langle \{ j^0(x), j^0(y) \} \rangle$
in that region from finite-temperature fluctuations of the plasma there,
leading, for example, to a contribution
\begin {equation}
   (\partial_{x^0} - D \grad_{\x}^2)
   (\partial_{y^0} - D \grad_{\y}^2)
   \left[
     \bigl\langle \{ j^0(x) , j^0(y) \} \bigr\rangle_{{\rm eq},T+\delta T}
   - \bigl\langle \{ j^0(x) , j^0(y) \} \bigr\rangle_{{\rm eq},T}
   \right]
\end {equation}
to the subtracted correlator (\ref{eq:jjsubtract}).
We will later see a Witten diagram in the gravity dual calculation
which naturally produce this sort of effect.
Fortunately, our refined measurement (\ref{eq:NewSigma}) would
eliminate this contribution since it only looks at charge density near
the light cone and so is not affected by effects on the charge
density far away, such as shown in fig.\ \ref{fig:sound}.

The definition (\ref{eq:NewSigma}) is an example of how to define
charge deposition, but it would be awkward and difficult to compute
$\tfrac12 \langle \{ \bar\Sigma_\Theta(\infty,x^3),
                     \bar\Sigma_\Theta(\infty,y^3) \rangle$
using this detailed formula in practice.
However, we will see that its effect is to simply
eliminate certain types of (gravity dual) diagrammatic contributions to the
charge-charge correlation
and to reproduce our naive
$\tfrac 12 \langle \{ \hat\Sigma_\Theta(\infty,x^3),
                      \hat\Sigma(\infty,y^3) \} \rangle$ for
the diagrammatic contributions which survive.  The upshot is that
in practice we will just compute
\begin {equation}
   \Bigl[
     \tfrac12
     \bigl\langle \{ \Sig(\infty,\x),
                     \Sig(\infty,\y) \}
     \bigr\rangle_{\rm jet}
   \Bigr]_{\rm relevant~diagrams}
\label {eq:computeme}
\end {equation}
We will see which are the relevant diagrammatic contributions
in sec.\ \ref{sec:a1dominance}.


\section {4-point functions and diagrammatic setup}
\label {sec:4point}

In this section, we see precisely which type of 4-point
equilibrium correlator we need to calculate and how it corresponds
to an appropriately causal real-time Witten diagram in the gravity
dual.  But first, it will be convenient to review the (r,a) notation
for organizing the classification of real-time correlators, based on
Schwinger-Keldysh formalism.


\subsection {Field theory review of \lowercase{(r,a)} notation}
\label {sec:rarules}

\begin {figure}
\begin {center}
  \includegraphics[scale=0.5]{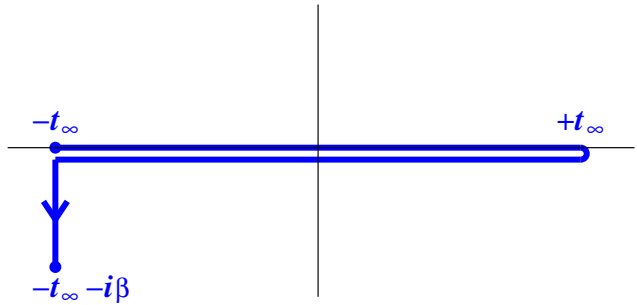}
  \caption{
     \label{fig:SKpath}
     The Schwinger-Keldysh contour in the complex time plane.
  }
\end {center}
\end {figure}

(r,a) notation
is nicely summarized in ref.\ \cite{WangHeinz}, and examples of its
application to calculations with gauge-gravity duality may be found
in refs.\ \cite{SonTeaney,HydroTails,CCT}.
One begins with Schwinger-Keldysh formalism for a system in
thermal equilibrium, where the
finite-temperature imaginary-time integration contour from
$0$ to $-i\beta$ is deformed into real time as shown in
fig.\ \ref{fig:SKpath}.%
\footnote{
  People sometimes find it useful to deform the contour so
  that there is a vertical segment that drops from
  $+t_\infty$ to $+t_\infty-i\sigma$ on the right-hand side
  of fig.\ \ref{fig:SKpath}, and then the vertical segment on
  the left-hand side drops from $-t_\infty-i\sigma$ to $-t_\infty-i\beta$.
  For instance, ref.\ \cite{HerzogSon} uses $\sigma=\beta/2$.
  But here we will stick with the traditional $\sigma=0$.
}
Operators may be placed on either the
upper or lower horizontal sections of the path, with operators
labeled type ``1'' or ``2'' correspondingly, and the
$n$-point functions are then path ordered.
For example, consider the
$n$-point correlator of a single field $O(x)$. Then
\begin {align}
   i^4 G_{12212}(x_1,x_2,x_3,x_4,x_5)
   &\equiv \bigl\langle {\cal T}_P [
        O_1(x_1) \, O_2(x_2) \, O_2(x_3) \, O_1(x_4) \, O_2(x_5)
    ] \bigr\rangle
\nonumber\\
   &\equiv
   \bigl\langle
      \bar{\cal T}[O(x_2) \, O(x_3) \, O(x_5)] \,
      {\cal T}[O(x_1) \, O(x_4)]
   \bigr\rangle ,
\end {align}
where ${\cal T}_P$ is Schwinger-Keldysh path ordering and
${\cal T}$ and $\bar{\cal T}$ represent ordinary time ordering and
reverse time ordering respectively.  For 2-point functions,
$G_{11}$ is the Feynman propagator $G_{\rm F}$;
$G_{22}$ is related to $G_{\rm F}^*$; and $G_{12}$ and
$G_{21}$ are the Wightman correlators $G_<$ and $G_>$.

The (r,a) notation consists of switching the basis of the labels
1 and 2 to%
\footnote{
  In the literature on the Schwinger-Keldysh formalism, $O_{\rm a}$
  and $O_{\rm r}$ are sometimes referred to, up to factor of 2 normalization
  conventions, as the ``quantum'' and ``classical'' fields
  $O_{\rm q}$ and $O_{\rm cl}$ respectively.  See, for example,
  ref.\ \cite{SonTeaney,Kamenev}.
}
\begin {subequations}
\label {eq:ranorm}
\begin {equation}
   O_{\rm a} \equiv O_1 - O_2 ,
   \qquad
   O_{\rm r} \equiv \tfrac12 (O_1 + O_2) .
\label {eq:ra}
\end {equation}
We will define the corresponding $n$-point Green functions by
\begin {equation}
   i^{n-1} G_{\alpha_1\cdots\alpha_n}(x_1,\cdots,x_n)
   = 2^{n_{\rm r}-1}
     \bigl\langle {\cal T}_P [
       O_{\alpha_1}(x_1) \, \cdots \, O_{\alpha_n}(x_n)
     ] \bigr\rangle ,
\end {equation}
\end {subequations}
where each $\alpha_i$ is either ``r'' or ``a'' and
$n_{\rm r}$ is the number of ``r'' indices.
Different authors use different factor of 2 conventions, and
ours are those of Wang and Heinz \cite{WangHeinz}.
For 2-point functions, $G_{\rm ar}(0,x)$ is the retarded
Green function $G_{\rm R}(x)$; $G_{\rm ra}(0,x)$ is the
advanced Green function $G_{\rm A}(x)$;
$iG_{\rm rr}(0,x)$ is the symmetric correlator
$\langle \{ O(0), O(x) \} \rangle = iG_> + iG_<$;
and $G_{\rm aa} = 0$.  An equilibrium relation that we will
find useful later on is the fluctuation-dissipation theorem,
\begin {equation}
   iG_{\rm rr}(\omega,\q)
   = - 2 \cth(\tfrac12\beta\omega) \, \Im G_{\rm R}(\omega,\q)
   = \cth(\tfrac12\beta\omega) \bigl[
       iG_{\rm R}(\omega,\q) - iG_{\rm A}(\omega,\q)
     \bigr]
\label {eq:fluctdisp}
\end {equation}
for bosonic correlators,%
\footnote{
   In addition to Kubo \cite{Kubo},
   see, for example, eq.\ (32) of Wang and Heinz \cite{WangHeinz}.
   Also $i G_>(\omega,\q) = -2[n(\omega)+1] \, \Im G_{\rm R}(\omega,\q)$ and
   $i G_<(\omega,\q) = -2n(\omega)\, \Im G_{\rm R}(\omega,\q)$,
   where
   $n(\omega) = 1/(e^{\beta\omega}-1) =
    \tfrac12 [\cth(\tfrac12\beta\omega)-1]$
   is the Bose distribution function.
   For a textbook reference, see, for example, eq. (7.6.41) of
   ref.\ \cite{ChaikinLubensky}, where their $2\chi''$ is the commutator
   $\langle[O(x),O(x')]\rangle = iG_{\rm R} - iG_{\rm A}$ and their $S$
   is the Wightman function $iG_> = \langle O(x)\,O(x') \rangle$.
   By taking $\omega \to -\omega$ one gets the corresponding
   relation for $G_<$, and then adding $G_>$ and $G_<$ gives
   $G_{\rm rr}$.
   As far as the case $\omega{=}0$ is concerned, we have implicitly
   assumed $\langle O \rangle = 0$, and so $O$ should be replaced by
   $O-\langle O \rangle$ if not.
}
which we have previously referenced in (\ref{eq:jjfluctdisp}).

Here is a useful property of correlators in (r,a) notation:
\begin {quote}
  {\it Rule 1}.
  An $n$-point function vanishes if the largest time
  $\max(t_1,\cdots,t_n)$
  is associated
  with an operator $O_{\rm a}$ that has an ``a'' label.
\end {quote}
This follows because, if $x^0$ is the largest time in the correlation,
then there is no difference between the operator ordering of
$\phi_1(x)$ and $\phi_2(x)$, and so $\phi_{\rm a}(x)$ defined by
(\ref{eq:ra}) vanishes.
Rule 1 has the following corollary:
\begin {quote}
  {\it Rule 2}.
  The $n$-point function $G_{\rm aa\cdots a}$
  with all ``a'' labels is zero.%
\footnote{
  This rule is equivalent to the largest-time diagrammar identity of Veltman
  \cite{Veltman}.
}
\end {quote}

\begin {figure}
\begin {center}
  \includegraphics[scale=0.4]{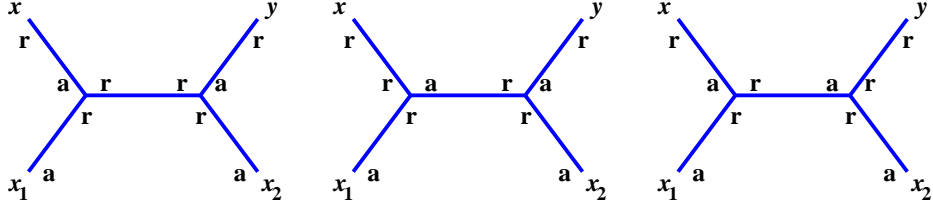}
  \caption{
     \label{fig:diag}
     The non-vanishing (r,a) assignments for the propagators in one
     of the diagrams that contribute to $G_{aarr}(x_1,x_2,x,y)$.
  }
\end {center}
\end {figure}

It is also useful to take note of a property of perturbation theory
in (r,a) formalism, even though the 4-dimensional field theory we
are interested in (strongly-coupled ${\cal N}{=}4$ super Yang-Mills)
is not weakly coupled.
In the Schwinger-Keldysh formalism, using the time contour of
fig.\ \ref{fig:SKpath}, the path integral schematically takes the form
\begin {equation}
   \int [{\cal D}\phi_1(x)] [{\cal D}\phi_2(x)]
   e^{i\int_x \bigl({\cal L}[\phi_1] - {\cal L}[\phi_2]\bigr)}
   e^{- S_{\rm E}[\phi_3]} ,
\end {equation}
where $\phi_1$ and $\phi_2$ live on the upper and lower horizontal
contour and $\phi_3$ lives on the vertical stub of the contour
at $\Re t{=}-t_\infty$.
In Hamiltonian language,
the $\phi_1$ and $\phi_2$ parts respectively generate the
evolution operators $e^{-i\int H \> dt}$ and $e^{+i \int H \> dt}$
that appear in the time
evolution $e^{+i\int H \> dt} O e^{-i\int H \> dt}$ of operators
that we wish to measure,
while the $\phi_3$ part generates the equilibrium density
matrix $e^{-\beta H}$.
Order by order in perturbation theory,
there are generically corrections both
to the evolution operators $e^{\pm i \int H \> dt}$ and to the
initial density matrix $e^{-\beta H}$.  For our application, only
corrections to the evolution will turn out to be important, and so
focus on the corresponding interaction terms
\begin {equation}
   \int_x \bigl( V[\phi_1] - V[\phi_2] \bigr)
   = \int_x \bigl( V[\phi_{\rm r} + \tfrac12 \phi_{\rm a}]
                 - V[\phi_{\rm r} - \tfrac12 \phi_{\rm a}] \bigr) .
\end {equation}
The integrand is odd in $\phi_{\rm a}$.  As a result we have the
following rule in (r,a) notation:
\begin {quote}
  {\it Rule 3}.
  Any interaction vertex (associated with time evolution)
  must have an odd number of ``a'' fields.
\end {quote}
As a simple example, consider the contribution of the diagram of
fig.\ \ref{fig:diag} to the 4-point correlator
$G_{\rm aarr}(x_1,x_2,x,y)$.  Applying Rules 1--3,
the figure shows all the non-vanishing (r,a) assignments for
this diagram's contribution to $G_{\rm aarr}$.

Of course, we will not be doing perturbation
theory in 4-dimensional field theory
for strongly-coupled ${\cal N}{=}4$ super Yang-Mills theory.
However, as others before us \cite{HydroTails,CCT},
we will extend the use of the
(r,a) notation to organize calculations in the 5-dimensional
gravitational dual theory.  In the gravitational theory, we are
doing perturbation theory.  Because the 5-dimensional
AdS$_5$-Schwarzschild
background is periodic in imaginary time with period $2\pi T$,
the methods of thermal field theory can be taken over to this
5-dimensional field theory problem \cite{GibbonsPerry}.

\begin {figure}
\begin {center}
  \includegraphics[scale=0.4]{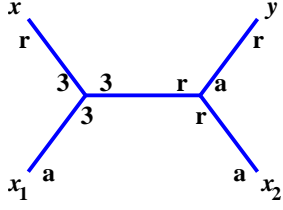}
  \caption{
     \label{fig:nodiag}
     An example of a diagram involving the fields $\phi_3$ on the
     vertical stub of the Schwinger-Keldysh time path of
     fig.\ \ref{fig:SKpath}.  Here the vertex marked with 3's
     represent a $(\phi_3)^3$ vertex on that stub.  This assignment
     (and all other assignments of this diagram that involve
     $\phi_3$ vertices)
     vanishes because $G_{3a} = 0$.
  }
\end {center}
\end {figure}

Finally, returning for a moment to ordinary perturbative thermal
quantum field theory, we should mention that there is a long,
involved story about whether and how, in the limit $t_\infty \to \infty$,
one may ignore vertices which correspond
to interaction terms such as $(\phi_3)^3$ that live on the
vertical stub of the contour in fig.\ \ref{fig:SKpath}.
(See Gelis \cite{vertical}.)  However, this is a subtlety
which may be blissfully ignored in many applications of interest.
For example, consider
a diagram like fig.\ \ref{fig:nodiag}, where
the ``3'' vertex represents a $(\phi_3)^3$ interaction along the
vertical stub of the contour in fig.\ \ref{fig:SKpath}.  This diagram
is zero because, by the same logic that
led to Rule 1, the propagator $G_{3a} = 0$.
At leading order, there are no non-vanishing diagrams for
$G_{\rm aarr}$ that involve corrections to the initial density matrix,
and so all the vertices in fig.\ \ref{fig:diag} just involve r's and
a's.
In the gravity dual, an analogous concern would arise if evaluating
a correlator whose causality structure allowed bulk vertices
to be inside the past or future horizon in fig.\ \ref{fig:witten} and
potentially touch the singularity, but this will not happen in our
application.  (When it does happen, ref.\ \cite{3point} shows how
to evaluate correlators in a way that
satisfies thermal equilibrium relations and does not require
consideration of points inside the horizon.  Alternatively,
one may use equilibrium relations \cite{WangHeinz}
to relate the correlator to another that avoids
the problem.  Or, as a matter of principle, one might use the
complexified AdS$_5$-Schwarzschild of Skenderis and van Rees
\cite{SvR1,SvR2}.)

\subsection{The desired 4-point correlator}

In our earlier work \cite{adsjet}, we needed an equilibrium 3-point
retarded correlator $G_{\rm aar}(x_1,x_2;x)$ to calculate
the response $\langle j^{0}(x) \rangle_{A_{\rm cl}}$ of
charge density to the creation of a jet, where each of the
``a'' subscripts in $G_{\rm aar}$ is associated with the source operator.
In this section, we
will see that we need the equilibrium 4-point correlator
$G_{\rm aarr}(x_1,x_2;x,y)$ to calculate the correlator
$\tfrac12 \langle \{ j^{0}(x) , j^{0}(y) \} \rangle_{A_{\rm cl}}$.
The appearance of ``r'' subscripts for both $x$ and $y$ in
$G_{\rm aarr}$ can be thought of as a generalization of the
equilibrium relation
$iG_{\rm rr} =
 \langle \{ O(x) , O(y) \} \rangle_{\rm eq}$
to the case of jets created by a source $A_{\rm cl}$.

Write the Hamiltonian of our system as
$H(t) = H_0 + \delta H(t)$,
where $\delta H(t)$ contains the small-amplitude source terms that we
introduce to create the jet and $H_0$ is everything else in the
full, interacting  Hamiltonian of the theory.
Take $\delta H(t)$ to be localized in time, as previously discussed,
and assume the system starts in equilibrium at $t \to -\infty$.
As reviewed in our earlier work \cite{adsjet}, the response of
an observable ${\cal O}$ is given by the following expansion in
$\delta H$:
\begin {equation}
   \Delta \langle {\cal O}(t) \rangle \equiv
   \langle {\cal O}(t) \rangle_H
   -
   \langle {\cal O} \rangle_{H_0}
   =
   \int dt_1 \> G_{\rm ar}(t_1;t)
   +
   \frac{1}{2!}
   \int dt_1 \> dt_2 \> G_{\rm aar}(t_1,t_2;t)
   +
   \cdots ,
\label {eq:fd}
\end {equation}
where the various $G_{\rm a\cdots ar}$ are the equilibrium
$n$-point retarded correlation functions, given in this case by
\begin {align}
   iG_{\rm ar}(t_1;t) &=
     \theta(t-t_1) \,
     \bigl\langle [{\cal O}(t),\delta H(t_1)] \bigr\rangle_{H_0} ,
\\
   i^2 G_{\rm aar}(t_1,t_2;t) &=
     \theta(t-t_2) \, \theta(t_2-t_1) \,
     \bigl\langle [[{\cal O}(t),\delta H(t_2)],\delta H(t_1)]
            \bigr\rangle_{H_0}
\nonumber\\ &
     +
     \theta(t-t_1) \, \theta(t_1-t_2) \,
     \bigl\langle [[{\cal O}(t),\delta H(t_1)],\delta H(t_2)]
            \bigr\rangle_{H_0} ,
\label {eq:G3generic}
\end {align}
etc.
Taking ${\cal O}$ to be $j^{(3)\mu}(x)$, and $\delta H$ from
(\ref{eq:SOURCE}) to be
$\int d^3x \> \j\cdot\A_{\rm cl}$, reproduces the formula (\ref{eq:jss})
that we reviewed earlier for $\langle j^{(3)\mu}(x) \rangle_{A_{\rm cl}}$.

Nothing changes in the derivation of the above formulas
if one replaces ${\cal O}(t)$ by
a product of operators ${\cal O}(t) \, {\cal O}(t')$
in the particular case that
$t$ and $t'$ are both after the time at which the source
has turned off (so that the source does not affect the time evolution
operator between $t$ and $t'$).
Such late-time observables are precisely what we're interesting in
for studying jet stopping.
We will focus in particular on the symmetric product
$\tfrac12 \{ {\cal O}(t) , {\cal O}(t') \}$ of measurement operators.
In the case of interest to jet quenching---a high-momentum source
$\delta H$ and low-momentum observables ${\cal O}$---the leading
contribution at post-source times to the $\delta H$ expansion will then be
\begin {equation}
   \tfrac12 \Delta \bigl\langle \{{\cal O}(t), {\cal O}(t')\} \bigr\rangle
   \simeq
   \frac{1}{2!}
   \int dt_1 \> dt_2 \> G(t_1,t_2;t,t')
\label {eq:fd2}
\end {equation}
with
\begin {align}
   i^2 G(t_1,t_2;t,t') &=
     \theta(t_2-t_1) \,
     \bigl\langle [[ \tfrac12 \{{\cal O}(t), {\cal O}(t')\},
                     \delta H(t_2)],\delta H(t_1)]
            \bigr\rangle_{H_0}
\nonumber\\ &
     + (t_1 \leftrightarrow t_2)
   .
\label {eq:GOOSS}
\end {align}
We have dropped factors such as $\theta(t-t_i)$ and $\theta(t'-t_i)$
because the formula (\ref{eq:fd2}) only applies in the case
of post-source times,
where $t$ and $t'$ are each greater than both $t_1$ and $t_2$.
In this case, it's easy to check that the (r,a) ordering
rules described in section \ref{sec:rarules} imply that
$G$ above is the same thing as $iG_{\rm aarr}/2$.
The factor of $i/2$ arises from our normalization convention
(\ref{eq:ranorm}) for (r,a) correlators.

Taking the same choices for ${\cal O}$ and $\delta H$ as before,
we get the following 4-point generalization of the 3-point
relation (\ref{eq:jss}):
\begin {equation}
  \tfrac12 \Delta
  \bigl\langle \{j^{(3)\mu}(x),j^{(3)\nu}(y)\} \bigr\rangle_{A_{\rm cl}}
  =
  \frac{i}{4}
  \int d^4x_1 \> d^4x_2 \>
  G_{\rm aarr}^{(ab33)\alpha\beta\mu\nu}(x_1,x_2;x,y) \,
  A^a_{\alpha,\rm cl}(x_1) \,
  A^b_{\beta,\rm cl}(x_2)
\label {eq:jjss}
\end {equation}
(for times $x^0,y^0$ after the source is turned off).


\subsection{Witten diagrams in the gravity dual}
\label {sec:diags}

\begin {figure}
\begin {center}
  \includegraphics[scale=0.5]{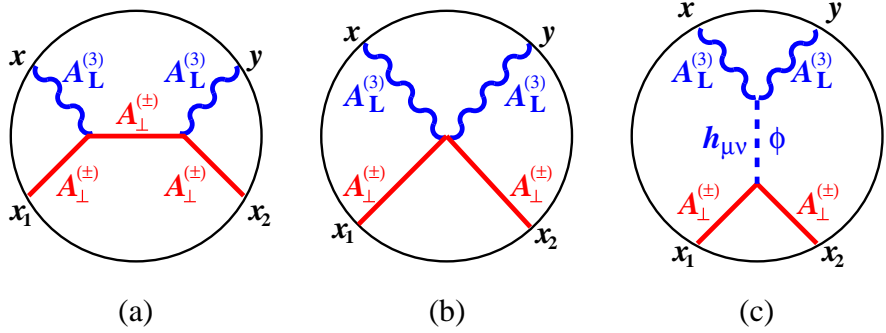}
  \caption{
     \label{fig:jjdiags}
     Witten diagrams contributing to the 4-point current correlator in
     our problem, with the sources at the bottom and the measurement
     points $x$ and $y$ at the top.  For (a), one should also add
     the diagram where $x$ and $y$ are interchanged.  The types of lines
     are described in the text.
  }
\end {center}

\end {figure}

Recall that we take our high-momentum source (\ref{eq:SOURCE}) to be
translation invariant in the transverse directions
($x^1, x^2$) and transversely polarized.
Restrict attention to measurements of correlators
of the charge density $j^0$, i.e. $\mu{=}\nu{=}0$ in
(\ref{eq:jjss}).
The Witten diagrams which contribute
to the
desired 4-point correlator in (\ref{eq:jjss}) are shown in
fig.\ \ref{fig:jjdiags}.
(Here, for the sake of pictorial simplicity, we have not bothered to
draw Penrose-diagram versions as in fig.\ \ref{fig:witten}.)
We are also not yet implementing the ``aarr'' prescription on
$G_{\rm aarr}$, which we will discuss shortly.
In these diagrams, straight solid lines represent propagators associated with
the transverse polarizations $A_\perp$ of the 5-dimensional gauge field.
These have high 4-momentum
and have index $\pm$ in the SU(2)
subgroup of the SU(4) R symmetry, corresponding to (\ref{eq:source}).
Wavy lines represent the $I{=}0$ and $3$
components of the gauge field $A_I$ (in $A_5{=}0$ gauge), which mix and which
we've abbreviated as $A_{\rm L}$ in the figure.
These wavy lines have low 4-momentum (since they are associated with our
late-time measurement of diffusion) and index 3 in the
SU(2) R-symmetry subgroup.  The dashed line can represent either
a graviton or a massive scalar $\phi$ discussed by Romans \cite{Romans}.

As seen from fig.\ \ref{fig:jjdiags}c, diagrams for the 4-point function
can contain other supergravity (SUGRA)
fields besides the ones dual to the boundary
operators.  The full set of fields in the SUGRA theory is rather
complicated, but the problem of enumerating the diagrams can be
simplified due to the fact that our sources all lie in an SU(2)
subgroup of the full SU(4) R symmetry.
L\"u, Pope, and Tran \cite{SU2U1} showed that the 5-dimensional
SU(4) gauge theory in the gravity dual can be consistently truncated
to a theory of just a SU(2)$\times$U(1) subgroup, and the Lagrangian
for ${\cal N}{=}4$ SUGRA in five dimensions with a gauged
SU(2)$\times$U(1) symmetry group is conveniently given by
Romans \cite{Romans}.  However, none of the details matter because
we will see later that only the diagram of fig.\ \ref{fig:jjdiags}a
will contribute to the physics we are interested in.

\begin {figure}
\begin {center}
  \includegraphics[scale=0.4]{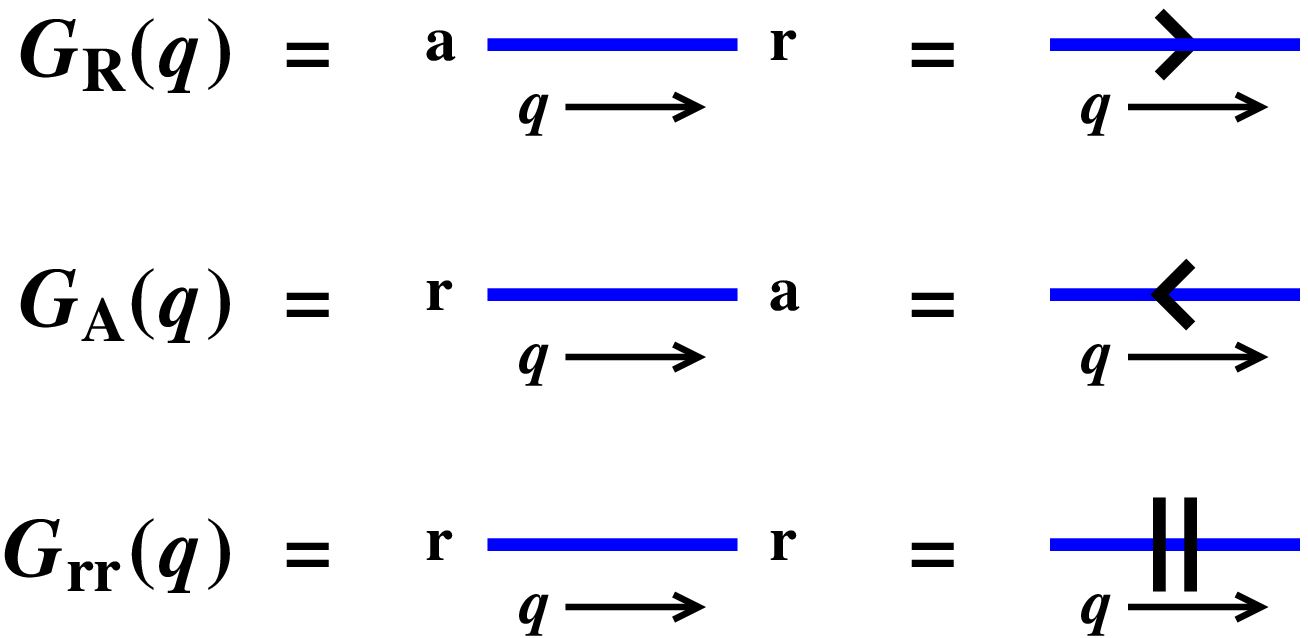}
  \caption{
     \label{fig:rakey}
     Our diagrammatic notation for propagators in the (r,a) formalism.
     The same convention is used for both bulk-to-bulk propagators
     $\Gbulk$ and bulk-to-boundary propagators ${\cal G}$.
  }
\end {center}
\end {figure}

\begin {figure}
\begin {center}
  \includegraphics[scale=0.4]{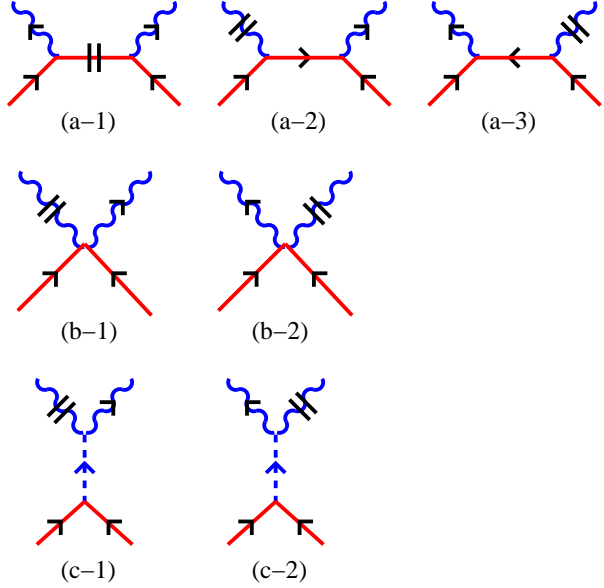}
  \caption{
     \label{fig:diagsra}
     Non-vanishing (r,a) assignments for
     Witten diagrams contributing to $G_{\rm aarr}$ in
     our problem.  The first row is simply a rewriting
     of fig.\ \ref{fig:diag} using the notation of fig.\ \ref{fig:rakey}.
  }
\end {center}
\end {figure}

\begin {figure}
\begin {center}
  \includegraphics[scale=0.4]{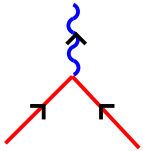}
  \caption{
     \label{fig:3pointra}
     Non-vanishing (r,a) assignment for
     the 3-point function $G_{\rm aar}$.
  }
\end {center}
\end {figure}

Now focus on $G_{\rm aarr}$ in particular.
In order to depict (r,a) assignments graphically, we will use
the notation defined in fig.\ \ref{fig:rakey}.
Using the rules
of section \ref{sec:rarules}, the (r,a) assignments for the
diagrams of fig.\ \ref{fig:jjdiags} which contribute to
$G_{\rm aarr}$ are shown in fig.\ \ref{fig:diagsra}.
For comparison, the diagram corresponding to the 3-point function
$G_{\rm aar}$ is shown in fig.\ \ref{fig:3pointra}.

\subsection {The dominance of diagram (a-1)}
\label {sec:a1dominance}

We will see now see that only diagram (a-1) potentially contributes to
the physics of interest to us.  The other diagrams, which all
involve a ${\cal G}_{\rm rr}$ on one of the two measurement legs (the wavy
lines), only
contribute to plasma-jet and other correlations of the sort discussed
and discarded in section \ref{sec:refine}.  In all of the diagrams,
we will make a low-momentum approximation for the 4-momenta $Q$ and
$Q'$ conjugate to the measurement points $x$ and $y$, as discussed at
the end of section \ref{sec:previous}.  That is, we will blur our eyes
and only concern ourselves with resolving measurements on distance
and time scales larger then $1/T$.

\begin {figure}
\begin {center}
  \includegraphics[scale=0.4]{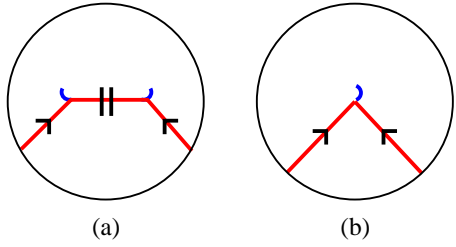}
  \caption{
     \label{fig:truncated}
     Versions of fig. \ref{fig:diagsra}(a-1) and fig.\
    \ref{fig:3pointra} with the measurement bulk-to-boundary
    propagators truncated by applications of the diffusion
    operator $\partial_t {-} D \grad^2$.  We have included a
    depiction of the
    boundary in these figures to emphasize that the truncated
    vertices lie in the bulk.
  }
\end {center}
\end {figure}

We start by discussing some of the properties of diagram (a-1).
The small 4-momentum limit of the gauge bulk-to-boundary propagator
${\cal G}^{\rm R}_{I\mu}$ is given in refs.\ \cite{SonStarinets,adsjet}.%
\footnote{
   See in particular eq. (4.7) of ref.\ \cite{adsjet}.
}
The details do not matter yet except that the general form is
\begin {equation}
   {\cal G}^{\rm R}_{I\mu}(u;\omega,\q)
   = \frac{f_{I\mu}(u;\omega,\q)}{i \omega - D\q^2} ,
\label {eq:GRsoft}
\end {equation}
where each $f_{I\mu}(u;\omega,\q)$ is polynomial in
$\omega$ and $\q$ (and so is local in 4-position space).
Above, the pole $(i\omega - D\q^2)^{-1}$
describes diffusion according to the diffusion equation.
If we now construct $\hat\Theta$ from $j^0$
for the measurement point,
by multiplying (\ref{eq:GRsoft})
by $i\omega-D\q^2$ as in the definition (\ref{eq:Theta}) of
$\hat\Theta$,
we get
\begin {equation}
   (\partial_t - D\grad^2) {\cal G}^{\rm R}_{I\mu}
   = \mbox{completely local in $x$} .
\end {equation}
That is, in 4-position space,
$(\partial_t - D\grad^2) {\cal G}^{\rm R}_{I\mu}$
connecting a bulk point $(x',u)$ to a boundary point $x$ only
has support at $x{=}x'$ (up to the $1/T$ smearing of resolution
intrinsic to our small-momentum approximation for the measurements).
So, once we apply the diffusion operators
$\partial_t{-}D\grad^2$ to both measurement points
as in (\ref{eq:TTdiffuse}), diagram (a-1) becomes a truncated
diagram depicted by fig.\ \ref{fig:truncated}a, with some derivatives
and possibly $u$-dependent factors that correspond to
$f_{I\mu}(u;\omega,\q)$ above acting on
the wavy stubs.  If we drew a corresponding diagram for
our earlier calculation \cite{adsjet} of
$\langle \hat\Theta(x) \rangle_{A_{\rm cl}}$, it would be
fig.\ \ref{fig:truncated}b.  In that
work, we found that the charge deposition only had
support for $x$ very close to the light cone,
where close means within $\max(L,T^{-1})$.  We will see the same
thing happen in this paper in the calculation corresponding to
fig.\ \ref{fig:truncated}a.  The underlying reason is that
the 5-dimensional wave created by the source, given by
a bulk-to-boundary propagator from a source point $x_i$
(represented by a straight line from the boundary in figs.\
\ref{fig:diagsra}--\ref{fig:truncated}) convolved
with the source wave packet $\Omega_L(x_i) e^{i\bar k\cdot x_i}$
on the boundary,
stays close to the $x^0{=}x^3$ 4-space light cone for all
times relevant to charge
deposition ($x^0 \lesssim E^{1/3}$).
We will leave the details of calculating
the contribution from diagram (a-1) until later, but this qualitative
point about the truncated diagram will be enough to understand what
happens with the other diagrams of fig.\ \ref{fig:diagsra}.

\begin {figure}
\begin {center}
  \includegraphics[scale=0.4]{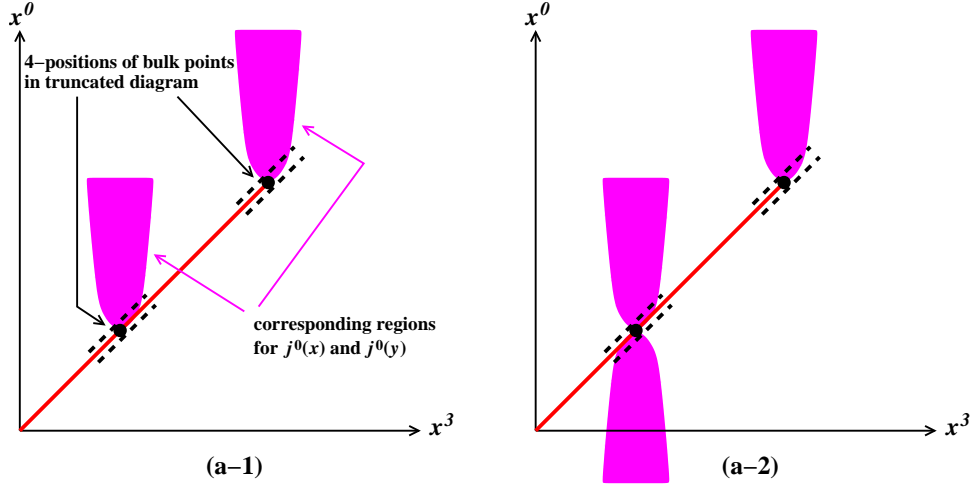}
  \caption{
     \label{fig:apicture}
     Regions of $x$ and $y$ that contribute to the current
     correlator $\tfrac12\langle\{j^0(x),j^0(y)\}\rangle$
     when the bulk points, which lie near the light cone,
     have 4-positions indicated by the black dots.
     The two diagrams correspond to the cases of diagrams
     (a-1) and (a-2) of fig.\ \ref{fig:diagsra}.
  }
\end {center}
\end {figure}

In diagram (a-1) of fig.\ \ref{fig:diagsra},
the charge-charge correlator
$\tfrac12 \langle \{j^0(x), j^0(y)\} \rangle_{A_{\rm cl}}$
was given by
deposition near the light cone,
represented by fig.\ \ref{fig:truncated}a,
propagated forward in time to the measurement points $x$ and $y$
by diffusive propagators, contained in the directed wavy lines
in fig.\ \ref{fig:diagsra}(a-1).  We depict this situation pictorially in
fig.\ \ref{fig:apicture}(a-1).
What happens if instead one of those
bulk-to-boundary propagators for $x$ or $y$
is a low-momentum ${\cal G}_{\rm rr}$ instead of a ${\cal G}_{\rm R}$,
as in diagram (a-2)?  Applying (\ref{eq:fluctdisp}) to the
gravity calculation,
\begin {align}
   {\cal G}_{\rm rr}(u;\omega,\q)
   &= \cth(\tfrac12\beta\omega) \bigl[
       {\cal G}_{\rm R}(u;\omega,\q) - {\cal G}_{\rm A}(u;\omega,\q)
     \bigr]
\nonumber\\
   &= \cth(\tfrac12\beta\omega) \bigl[
       {\cal G}_{\rm R}(u;\omega,\q) - {\cal G}_{\rm R}(u;-\omega,-\q)
     \bigr] .
\label {eq:calfluctdisp}
\end {align}
As a result, ${\cal G}_{\rm rr}(u;q)$ above is even under inversion $q\to-q$
of the 4-momentum.  Fourier transforming in 4-space, that means that
${\cal G}_{\rm rr}(x',u;x)$ is even in the difference $x-x'$ of the boundary
and bulk 4-positions.
We depict this situation pictorially in fig.\ \ref{fig:apicture}(a-2),
where the point associated with ${\cal G}_{\rm rr}$ has a
backward-time diffusive region, corresponding to the second term
in (\ref{eq:calfluctdisp}), that is $PT$-symmetric with the
forward-time diffusive region.
As a result, if we calculate the refined time-integrated
charge-deposition
operator $\bar\Sigma_\Theta$ of (\ref{eq:NewSigma}) at a point on the light
cone, the contributions from subtracting the integrals of $j^0$ over the
two dashed hyper-plane segments in the figure will cancel.  And so
diagram (a-2) will not contribute to our refined measure
$\tfrac12 \langle \{ \bar\Sigma_\Theta(\infty,x^3) ,
                     \bar\Sigma_\Theta(\infty,y^3) \} \rangle$
of charge deposition correlations.  The same conclusion holds for
diagram (a-3) and, by similar logic, for (b-1) and (b-2).

In what follows, we will abbreviate $\bar\Sigma_\Theta(\infty,x^3)$
as $\bar\Sigma(x^3)$.

\begin {figure}
\begin {center}
  \includegraphics[scale=0.4]{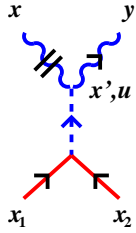}
  \caption{
     \label{fig:c1diag}
     Labeling of points for discussing diagram (c-1) of
     fig.\ \ref{fig:diagsra}.
  }
\end {center}
\end {figure}

\begin {figure}
\begin {center}
  \includegraphics[scale=0.4]{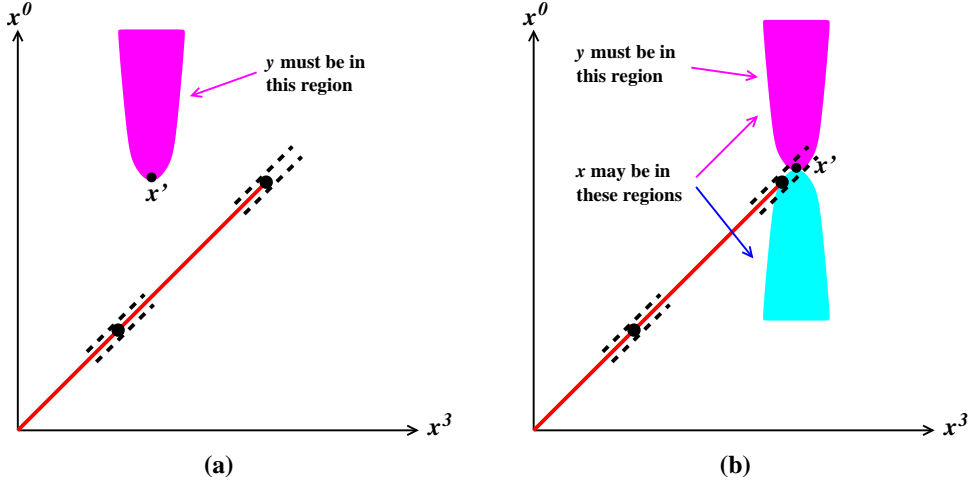}
  \caption{
     \label{fig:c1picture}
     Restrictions on where the measurement points $x$ and $y$ can
     be for diagram (c-1) in the cases where the bulk point
     $x'$ of fig.\ \ref{fig:c1diag} is (a) far from or (b) near
     the $x^0 \simeq x^3$ light cone.
     In the first case, $x'$ cannot be below the lightcone because
     of causality constraints.
  }
\end {center}
\end {figure}

Now we come to the (symmetrically related) diagrams (c-1) and (c-2) of fig.\
\ref{fig:diagsra}.  Focus on (c-1) for the sake of specificity and
let $(x',u)$ be the location of the top-most bulk vertex, as
indicated in fig.\ \ref{fig:c1diag}.  
Because of the flow of causality (indicated by the arrows) from the
source points $x_1$ and $x_2$, the bulk point $(x',u)$ must lie in
the 5-dimensional causal future of $x_1$ and $x_2$, and the measurement
point $y$ must lie in the future of that.
This requires $x'$ to
be in the 4-dimensional causal future of the source and $y$ to be
in the causal future of $x'$.  Now consider two cases: $x'$ is (i) far
from or (ii) near the $x^0 \simeq x^3$ lightcone, as in
figs.\ \ref{fig:c1picture}a and b respectively.  In the first case,
it is causally impossible for the measurement point $y$ to be near
the light cone, and so there can be no contribution to the
refined correlator
$\tfrac12 \langle \{ \bar\Sigma(x^3),
                     \bar\Sigma(y^3) \} \rangle$,
which is defined to
only involve measurements near the light cone.  In the second case, there
can be a contribution but only if the measurement points $x$ and $y$ are
very close to $x'$ and so to each other.
That's because $y$ is connected to
$(x',u)$ by a low-momentum ${\cal G}_{\rm R}$ and so must lie in the
forward-time diffusive region relative to $x'$, as depicted by the
magenta region in fig.\ \ref{fig:c1picture}b, while $x$ is connected by
a low-momentum ${\cal G}_{\rm rr}$ and so may lie in either the
forward-time or backward-time regions, as depicted by the magenta and
blue regions of fig.\ \ref{fig:c1picture}b.  If both are to be measured
close to light cone to determine
$\tfrac12 \langle \{ \bar\Sigma(x^3),
                     \bar\Sigma(y^3) \} \rangle$,
then all the points are close to each other, and so diagrams
(c-1) and (c-2) only contribute to very localized correlations.
Our goal, however, is to understand whether
$\tfrac12 \langle \{ \bar\Sigma(x^3),
                     \bar\Sigma(y^3) \} \rangle$
has support at relatively large separations.  To answer that question,
we may therefore ignore (c-1) and (c-2).

The low-momentum graviton propagator has a sound pole.%
\footnote{
   See, for example, ref.\ \cite{sound}, which derives
   sound poles in the bulk-to-boundary graviton
   propagator and then in 2-point boundary correlators.
   We will discuss later the relation of
   bulk-to-boundary and bulk-to-bulk propagators at finite
   temperature.
}
Though not necessary for the above argument, we note in passing that
for a graviton internal propagator in (c-1), the picture in
fig.\ \ref{fig:c1picture}a therefore plausibly has some relation
to the type of physics discussed earlier in fig.\ \ref{fig:sound}.

The upshot of this section is that we may compute the refined
correlator
$\tfrac12 \langle \{ \bar\Sigma(x^3),
                     \bar\Sigma(y^3) \} \rangle$,
of section \ref{sec:refine} with only diagram (a-1).  But the causality
properties of diagram (a-1) are such that the reasons for carefully defining
the refined correlator no longer apply.  So, restricting attention to
(a-1), there is no reason not to compute the original, simpler correlator
$\tfrac12 \langle \{ \hat\Theta(x),
                     \hat\Theta(y) \} \rangle$,
or its time-integrated
version
$\tfrac12 \langle \{ \Sig(x^0{=}\infty,\x),
                  \Sig(y^0{=}\infty,\y)\} \rangle$.
That is, we will compute (\ref{eq:computeme}) with the
``relevant diagrams'' being (a-1).


\section {The bulk-to-bulk correlator \boldmath$\Grrbulk$}
\label {sec:Grrbulk}

In order to evaluate diagram (a-1) of fig.\ \ref{fig:diagsra},
we will need the
bulk-to-bulk correlator $\Grrbulk$ associated with the internal line
of that diagram.  (The superscript is a reminder that this is the
5-dimensional bulk correlator and not the 4-dimensional boundary
correlator.)
When evaluating correlators of the time-integrated charge
deposition $\Sig(x^0{=}\infty,\x)$, we will turn out to need only
the case where the two bulk points in diagram (a-1) are at very large
times.
At very
large times, the 5-dimensional excitation created by the source has fallen
in the 5th dimension, so that it is localized very near the horizon.
In consequence, we will really only need the bulk-to-bulk correlator
$\Grrbulk$ between near-horizon points.

In section \ref{sec:Grrgeneral} below, we give a general discussion
of how bulk-to-bulk propagators such as $\Grrbulk$ can be computed in terms
of more familiar bulk-to-boundary propagators ${\cal G}$.
We then specialize in section \ref{sec:Grrhorizon} to the case where
both bulk points lie very near the horizon.


\subsection {General case}
\label{sec:Grrgeneral}

For diagram (a-1) of fig.\ \ref{fig:diagsra},
we are interested in bulk-to-bulk propagators for a
transversely-polarized bulk gauge field $A_\perp$ that is dual
to the transversely-polarized R-charge current operator $j_\perp$.
The generalization to other cases is relatively straightforward.
In particular, in this section we will simultaneously treat the
case of a bulk scalar field, which does not require adding any
additional indices or other complications of notation and
which is relevant to our discussion of
$e_{\rm jet}\gg 1$ in Appendix \ref{app:Rcharge}.

\subsubsection {Bulk-to-bulk $\GRbulk$}

We start by looking at the retarded bulk-to-bulk propagator
$\GRbulk$ from one bulk point $X_1=(x_1,u_1)$ to another $X_2=(x_2,u_2)$.
Note that $x_1$ and $x_2$ label 4-positions of bulk points in this
discussion and are not the source points we have labeled $x_1$ and $x_2$
previously.  Let $q$ be the Fourier conjugate of $x_2{-}x_1$.
We will verify in a moment that $\GRbulk$ can be written in terms of
bulk-to-boundary propagators as
\begin {equation}
   \GRbulk(u_1;u_2,q) \equiv \Gbulk_{\rm ar}(u_1;u_2,q) =
   \frac{{\cal N}_{\rm G}}{W_{\rm RA}(q)} \,
   \calGR(u_>;q) \left[ \calGR(u_<;q) - \calGA(u_<;q) \right] ,
\label {eq:GRbulk}
\end {equation}
where
$W_{\rm RA}(q)$ is the Wronskian (to be made explicit in a moment),
$u_< \equiv \min(u_1,u_2)$, and $u_> \equiv \max(u_1,u_2)$.
The overall normalization ${\cal N}_{\rm G}$ comes from the
normalization of the kinetic term in the 5-dimensional Lagrangian.
For gauge fields, we take
${\cal L}_{\rm kin} = -\tfrac14 (\gSG^2 R)^{-1} F^{IJa} F^a_{IJ}$
and correspondingly
\begin {equation}
   {\cal N}_{\rm G} = \gSG^2 R ,
\end {equation}
where $R$ is the AdS radius and $\gSG = 4\pi/N_{\rm c}$
\cite{Witten,Freedman}.  For scalars,
${\cal L}_{\rm kin}
 = - (\gSG^2 R^3)^{-1} (\partial_I \phi) g^{IJ} (\partial_J \phi)$
and ${\cal N}_{\rm G} = \gSG^2 R^3/2$.

Let's now check that the propagator (\ref{eq:GRbulk}) satisfies the
equation of motion with a source term at $(x_1,u_1)$:
\begin {equation}
   \frac{1}{{\cal N}_{\rm G}}
   \Bigl[
     \partial_I \bigl(
        \sqrt{-g} \, g^{IJ} \lambda \, \partial_J
     \bigr)
     - \sqrt{-g} \, m^2
   \Bigr] \GRbulk(X_1;X)
   = \delta^{(5)}(X-X_1) ,
\end {equation}
where
\begin {equation}
   \lambda \equiv
   \begin {cases}
     g^{\perp\perp} , & \mbox{transverse gauge field $A_\perp$}; \\
     1              , & \mbox{scalar field $\phi$} ,
   \end {cases}
\end {equation}
and $m=0$ in the gauge field case.  In 4-momentum space, this is
\begin {equation}
   \frac{1}{{\cal N}_{\rm G}}
   \Bigl[
     \partial_5 \bigl(
        \sqrt{-g} \, g^{55} \lambda \, \partial_5
     \bigr)
     - \sqrt{-g} \, (\lambda q_\mu g^{\mu\nu} q_\nu + m^2)
   \Bigr] \GRbulk(u_1;u,q)
   = \delta(u-u_1) .
\label {eq:EOMsource}
\end {equation}
From the fact that the bulk-to-boundary propagators satisfy the
homogeneous equation
\begin {equation}
   \frac{1}{{\cal N}_{\rm G}}
   \Bigl[
     \partial_5 \bigl(
        \sqrt{-g} \, g^{55} \lambda \, \partial_5
     \bigr)
     - \sqrt{-g} \, (\lambda q_\mu g^{\mu\nu} q_\nu + m^2)
   \Bigr] {\cal G}(u;q)
   = 0 ,
\label {eq:EOMhomo}
\end {equation}
one can verify that (\ref{eq:GRbulk}) satisfies (\ref{eq:EOMsource}) provided
$W_{\rm RA}(q)$ is the Wronskian
\begin {equation}
   W_{\rm RA} = \sqrt{-g}\,g^{55} \lambda \,
   (\calGR \tensor\partial_5 \calGA) ,
\label {eq:WRA}
\end {equation}
which is independent of $u$.
Here, $\tensor\partial$ is defined by
$f\tensor\partial g \equiv f\partial g - (\partial f) g$.

For the bulk-to-boundary propagator ${\cal G}(u;q)$, our
convention is that $q$ is the momentum conjugate to the boundary point.
For the bulk-to-bulk correlator $\GRbulk(u_1;u_2,q)$,
the convention is that $q$ is associated with
the bulk point at $u_2$, which we take to be the later-time point
in defining the retarded correlator.
Eq.\ (\ref{eq:GRbulk}) gives a {\it retarded}\/ bulk-to-bulk correlator because
it is analytic in the upper-half $q^0$ plane since (i) $\calGR(u;q)$
has this property
and (ii) the poles in $\calGA(u_<;q)$ cancel the corresponding poles in the
denominator $W_{\rm RA}$.

Finally, bulk-to-bulk propagators should satisfy the correct boundary
condition at the boundary, which is that the boundary is fixed and so
fluctuation fields (such as $\phi$ and $A_\perp$) should vanish there
in this context.  That is, $\Gbulk(u_1;u_2,q)$ should vanish for
$u_1$ or $u_2$ zero.  This is easy to see for (\ref{eq:GRbulk}) in
the case of $A_\perp$ or massless scalar fields, since the
bulk-to-boundary propagators are then normalized to 1 at the
boundary, so that $\calGA(0;q)-\calGR(0;q) = 0$.
Eq.\ (\ref{eq:GRbulk}) also satisfies the boundary condition in
the massive scalar case, but there one must take care to
regulate the boundary.


\subsubsection {Bulk-to-bulk $\GAbulk$}

The advanced bulk-to-bulk propagator is similarly
\begin {equation}
   \GAbulk(u_1;u_2,q) \equiv \Gbulk_{\rm ra}(u_1;u_2,q) =
   \frac{{\cal N}_{\rm G}}{W_{\rm RA}(q)} \,
   \calGA(u_>;q) \left[ \calGR(u_<;q) - \calGA(u_<;q) \right] .
\label {eq:GAbulk}
\end {equation}


\subsubsection {Bulk-to-bulk $\Grrbulk$}

Applying (\ref{eq:fluctdisp}) then produces
\begin {align}
   \Grrbulk(u_1;u_2,q)
   &=
   \cth(\tfrac12\beta\omega)
   \bigl[ \GRbulk(u;q) - \GAbulk(u;q) \bigr]
\nonumber\\
   &=
   \cth(\tfrac12\beta\omega) \,
   \frac{{\cal N}_{\rm G}}{W_{\rm RA}(q)} \,
   \left[ \calGR(u_1;q) - \calGA(u_1;q) \right]
   \left[ \calGR(u_2;q) - \calGA(u_2;q) \right] .
\label {eq:Grrbulk}
\end {align}


\subsection {Near-horizon case}
\label{sec:Grrhorizon}

Here and throughout the rest of this paper, we will work in units
where $2\pi T{=}1$ and specialize to
the choice of 5th dimension coordinate $u$ with metric (\ref{eq:metric}).

\subsubsection{Near-horizon bulk-to-boundary propagator}

Near the horizon, the general solution to the homogeneous equation of motion
(\ref{eq:EOMhomo}) with the boundary conditions appropriate to
$\calGR$ can be expanded in powers of $1{-}u$ as
\begin {equation}
   \calGR(u;q) = b(q) \, (1-u)^{-i\omega/2} [1 + O(1-u)] .
\label {eq:calGRnearu}
\end {equation}
We will not need the exact solution away from the horizon.%
\footnote{
  \label{foot:heun}
  As an example, for the case of $A_\perp$, the exact solution
  is given by replacing $[1+O(1-u)]$ in (\ref{eq:calGRnearu})
  by
  $ [\tfrac12 (1+u)]^{\omega/2}
    \Hl\bigl(
      2,
      q^2 - \tfrac{i}{2}\omega^2 + \tfrac{1-i}{2}\omega;
      \tfrac{1-i}{2}\omega,
      \tfrac{1-i}{2}\omega + 1,
      1-i\omega,
      0;
      1-u
   \bigr)$,
  where $\Hl$ is a Heun function as defined in ref.\ \cite{Heun}.
}
All of
its complexity is summarized here in the factor $b(q)$, which is
determined by the normalization condition for $\calGR$ on the
boundary (e.g.\ ${\calGR}(0;q)=1$ for $A_\perp$ and massless scalar fields).
The answer to our question of whether jet charge deposition is
correlated over relatively large distances will turn out not to depend
on the details of $b(q)$ --- all that will matter are some simple analytic
properties of $b(q)$.
Since $\calGR$ is a retarded propagator, $b(q)$ must be analytic in
the upper-half complex frequency plane.  Because
$[\calGR(u;-q)]^* = \calGR(u;q)$ for real $q$,
(\ref{eq:calGRnearu}) also gives the important
relation
\begin {equation}
    b(-q) = b^*(q)
    \qquad
    \mbox{for real $q$}.
\label {eq:bconj}
\end {equation}

It will be convenient to change variables from $u$ to
\begin {equation}
  \tau \simeq -\tfrac12 \ln(1-u) .
\label {eq:tau}
\end {equation}
The $\simeq$ indicates that the details of the definition will not
matter with regard to sub-leading corrections as $u \to 1$.
Physically, $\tau$ represents the time (as measured by an asymptotic
observer) for the a wave in the bulk to propagate from near the boundary to the
near-horizon position $u \simeq 1$, and details such as initial
conditions or phase vs.\ group velocity
only affect sub-logarithmic corrections to
(\ref{eq:tau}).  In terms of $\tau$, (\ref{eq:calGRnearu}) becomes
\begin{equation}
   \calGRup_\perp(u;q) =
   b(q)\,e^{i\omega\tau} \bigl[ 1 + O(e^{-2\tau}) \bigr] .
\label {eq:calGRnear}
\end {equation}


\subsubsection{Near-horizon bulk-to-bulk $\GRbulk$}

We now turn to $\GRbulk$ as given by (\ref{eq:GRbulk}).  For simplicity
of presentation, we will specialize here to the case of the $A_\perp$
propagator and quote the scalar case at the end.
We first need the Wronskian (\ref{eq:WRA}).
Since the Wronskian is $u$-independent, we can evaluate (\ref{eq:WRA})
at any $u$, and it is most convenient to evaluate it in the horizon
limit $u\to 1$.  Using (\ref{eq:calGRnearu}) and its conjugate
\begin{equation}
   \calGA(u;q) =
   \calGR(u;-q) =
   b(-q)\,(1-u)^{i\omega/2} \bigl[ 1 + O(1-u) \bigr] ,
\end {equation}
together with
\begin {equation}
   \sqrt{-g} \, g^{55} g^{\perp\perp}
   = \tfrac12 \, Rf
   \to R(1-u) ,
\end {equation}
we get
\begin {equation}
   W_{\rm RA} = -i R \omega \, b(q) \, b(-q) ,
\label {eq:WRAT}
\end {equation}
which for real $q$ is $W_{\rm RA} = -i R \omega |b(q)|^2$.
Combining (\ref{eq:WRAT}) with (\ref{eq:GRbulk}) and (\ref{eq:calGRnear}), the
near-horizon limit of the retarded bulk-to-bulk propagator is
\begin {align}
   i\GRbulk(\tau_1;\tau_2,q) &\simeq
   \frac{\gSG^2}{\omega} \,
   e^{i\omega \tau_>}
   \left[
      - e^{i\delta(q)} e^{i\omega \tau_<} + e^{-i\omega\tau_<}
   \right]
\nonumber\\
   &=
   \frac{\gSG^2}{\omega}
   \left[
      - e^{i\delta(q)} e^{i\omega (\tau_2+\tau_1)} + e^{i\omega|\tau_2-\tau_1|}
   \right] ,
\label {eq:GRnear}
\end {align}
where
\begin {equation}
  e^{i\delta(q)} \equiv \frac{b(q)}{b(-q)} \,.
\end {equation}
For real $q$, the phase $\delta(q)$ is real.
At poles of $\GRbulk$ in the lower-half complex frequency plane,
$\delta$ becomes $-i\infty$.
The right-hand side of (\ref{eq:GRnear}) is finite as $\omega\to0$
because%
\footnote{
   Presumably there is some simple way to understand
   (\ref{eq:deltalim}),
   but we just checked it numerically.
}
\begin {equation}
   \delta(\omega,\q) \to 0 \qquad\mbox{linearly as}\qquad \omega\to 0 .
\label {eq:deltalim}
\end {equation}
For scalar fields, the result is a factor of 2 larger than
(\ref{eq:GRnear}).

\begin {figure}
\begin {center}
  \includegraphics[scale=0.4]{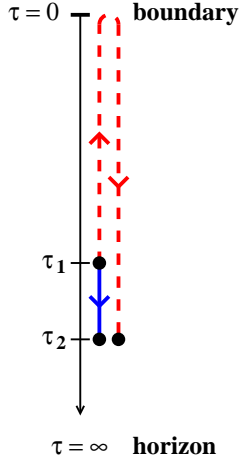}
  \caption{
     \label{fig:GRbounce}
     Schematic representation of the two contributions to the
     near-horizon bulk-to-bulk retarded propagator (\ref{eq:GRnear}).
  }
\end {center}
\end {figure}

To understand the physical interpretation of the two
terms on the right-hand side of (\ref{eq:GRnear}), consider
the metric (\ref{eq:metric}) in the near-horizon limit,
\begin {equation}
   ds^2 \simeq
   \frac{R^2}{4} \left[ - f \, dt^2 + d\x^2 + f \, d\tau^2 \right]
\end {equation}
with $f \simeq 2(1-u) \simeq 2e^{-2\tau}$.  Note that points
with different $\x$ are far apart compared to points with different
$t$ or $\tau$.  Because of this, the $\x$ motion decouples.
More concretely, the equation of motion (\ref{eq:EOMsource})
satisfied by the $A_\perp$ propagator is
\begin {equation}
   (\partial_\tau^2 + \omega^2) \, \GRbulk(\tau_1;\tau,q) \simeq
   2\gSG^2 \, \delta(\tau - \tau_1)
\label {eq:EOMnear}
\end {equation}
in the near-horizon limit for fixed $q$.  This looks similar to
a flat-space problem in one space-time dimension, where the Green function
would be proportional to $(i\omega)^{-1} e^{i \omega |\Delta\tau|}$.
Compare to (\ref{eq:GRnear}).
The second term on the right-hand side of (\ref{eq:GRnear})
corresponds to a signal that travels directly from $\tau_1$
to $\tau_2$, corresponding to a distance $|\tau_2-\tau_1|$ in $\tau$.
The first term, in contrast, corresponds to a signal
that travels from $\tau_1$ to the boundary $\tau{=}0$ and then
reflects back to $\tau_2$, traveling a total $\tau$-distance
proportional to $\tau_2+\tau_1$.  $\delta(q)$ represents the
phase shift that the wave picks up traveling through the region far
from the horizon, where the dynamics becomes more complicated than
the simple asymptotic behavior (\ref{eq:EOMnear}).  The relative
minus sign between the two terms is because the wave
flips when it reflects from the boundary due to the boundary condition
that the Green function vanish
at the horizon.  The two contributions to the right-hand side of
(\ref{eq:GRnear})
are depicted schematically in fig.\ \ref{fig:GRbounce}.

\subsubsection {Near-horizon bulk-to-bulk $\Grrbulk$}

Applying the same results to the formula (\ref{eq:Grrbulk}) for
$\Grrbulk$ gives, in units where $2\pi T=1$,
\begin {align}
  i\Grrbulk(u_1;u_2,q) &\simeq
  \gSG^2 \, \frac{\cth(\pi\omega)}{\omega}
  \left(
    e^{i\omega\tau_1} - e^{-i\delta(q)} e^{-i\omega\tau_1}
  \right)
  \left(
    e^{-i\omega\tau_2} - e^{i\delta(q)} e^{i\omega\tau_2}
  \right)
\nonumber\\
  &= 2 \gSG^2 \, \frac{\cth(\pi\omega)}{\omega}
  \left[
       - \cos\Bigl(\omega(\tau_2+\tau_1) + \delta(q)\Bigr)
       + \cos\Bigl(\omega(\tau_2-\tau_1)\Bigr)
  \right] .
\label {eq:Grrnear}
\end {align}
Note that the $\omega{\to}0$ limit is finite because of
(\ref{eq:deltalim}).

The second term in (\ref{eq:Grrnear}) is independent of $\q$ and so
is a $\delta$-function $\delta^{(3)}(\x_1-\x_2)$ in 3-position space.
This locality is a consequence of the fact that $\x$ motion is
suppressed near the horizon.
In contrast, the first term does depend on $\q$ through the phase $\delta(q)$
and so can produce non-local correlations in $\x$.  This non-locality
arises because the phase $\delta$ appears in contributions like the
dashed line in fig.\ \ref{fig:GRbounce}, corresponding
to propagation away from
the horizon, so that 3-space dynamics are then nontrivial, bouncing off of the
boundary, and then propagating back to near the horizon.
We will see that only the second term in (\ref{eq:Grrnear}) contributes
significantly to charge deposition correlations for high-energy jets,
and the locality of that term will be responsible for the
locality of the charge deposition correlation.

In Appendix \ref{app:CCT}, we show how to relate our result
(\ref{eq:Grrnear}) to expressions derived for bulk-to-bulk $G_{\rm rr}$
by Caron-Huot, Chesler, and Teaney \cite{CCT}
in the context of fluctuations of a classical string.


\section {Calculating the correlator}
\label {sec:calculate}

\subsection {Starting formula}

We are now finally ready to calculate the contribution of diagram
(a-1) of fig.\ \ref{fig:diagsra} to our correlator.
In our application, only the standard Yang-Mills 3-point vertices
contribute to this diagram: the contributions from the Chern-Simons
term in the SUGRA Lagrangian vanish because the SU(4)
group factor $d^{\pm3e} \equiv 2 \tr[\{T^\pm,T^3\}T^e]$ that
would be associated with a Chern-Simons 3-point vertex vanishes.

To save space, we will henceforth abbreviate 
$\langle \tfrac12 \{ j^{(3)\mu}(x) , j^{(3)\nu}(y) \} \rangle$
as $\langle \tfrac12 \{ j^\mu,j^\nu\} \rangle$.
Plugging the source (\ref{eq:source}) into the relationship
(\ref{eq:jjss}) between $\langle \tfrac12 \{ j^\mu,j^\nu\} \rangle$ and
the 4-point equilibrium correlator $G_{\rm aarr}$, and expressing
$G_{\rm aarr}$ in momentum space, gives
\begin {align}
  \Delta\langle \tfrac12 \{ j^\mu,j^\nu\} \rangle
  &=
  i \, \frac{\Aamp^2}{2}
  \int_{QQ'Q_1Q_2} \pol_\alpha \pol_\beta \,
  G_{\rm aarr}^{({-}{+}33)\alpha\beta\mu\nu}(Q_1,Q_2;Q,Q') \,
  \tilde\Lambda_L^*(Q_1-\bar k) \,
  \tilde\Lambda_L^*(Q_2+\bar k) \,
\nonumber\\ & \hspace{10em} \times
  e^{iQ\cdot x}
  e^{iQ'\cdot y}
  (2\pi)^4 \delta^{(4)}(Q{+}Q'{+}Q_1{+}Q_2) .
\label {eq:basic2}
\end {align}

\begin {figure}
\begin {center}
  \includegraphics[scale=0.5]{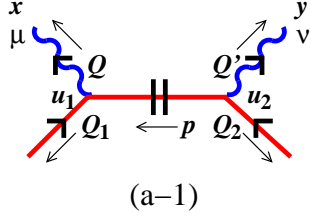}
  \caption{
     \label{fig:a1momenta}
     Naming conventions for momenta assignments to diagram (a-1).
  }
\end {center}
\end {figure}

Labeling momenta as in fig.\ \ref{fig:a1momenta},
diagram (a-1) plus its permutations gives
\begin {align}
   \left[
    \pol_\alpha \pol_\beta (G_{\rm aarr})^{({-}{+}33)\alpha\beta\mu\nu}
   \right]_{\rm(a-1)}
   &= i f^{{-}e3} f^{e{+}3}
\nonumber\\ & \quad \times
   \frac{i(-p-Q_1)_\rho}{\gSG^2 R}
   \int du_1 \> (\sqrt{-g} g^{\perp\perp} g^{\rho\sigma})_1 \,
   \calGAup_\perp(u_1;Q_1) \, {\cal G}^{{\rm R}\mu}_\sigma(u_1;Q)
\nonumber\\ & \quad \times
   \frac{i(Q_2-p)_\xi}{\gSG^2 R}
   \int du_2 \> (\sqrt{-g} g^{\perp\perp} g^{\xi\tau})_2 \,
   \calGAup_\perp(u_2;Q_2) \, {\cal G}^{{\rm R}\nu}_\tau(u_2;Q')
\nonumber\\ & \quad \times
   i \Grrperpbulk(u_1;u_2,-p) \biggl|_{p=Q_1+Q=-(Q_2+Q')}
\nonumber\\ &
   + \{Q,\mu\leftrightarrow Q',\nu\} .
\label {eq:Gaarr1}
\end {align}
The group factors are $f^{-e3}f^{e+3} = -2$.
In Appendix \ref{app:Tzero}, we show that these formulas give a
simple and easily predictable result for
$\Delta\langle \tfrac12 \{ j^\mu,j^\nu\} \rangle$ in the $T{=}0$
case.
But here we focus on finite temperature and the contribution of
diagram (a-1) to
$\tfrac12 \langle\{ \Sig(x^0{=}\infty,\x),
                    \Sig(y^0{=}\infty,\y)\}\rangle$,
which we will abbreviate as
$\tfrac12 \langle\{ \Sig,  \Sig \rangle\}$
in the following.

From the definitions (\ref{eq:Sigma}) and (\ref{eq:Theta}) of
$\Sig$ and $\hat\Theta$,
replacing $j^\mu$ and $j^\nu$ by $\Sig$ in (\ref{eq:basic2})
is equivalent to replacing
\begin {equation}
  {\cal G}^{{\rm R}\mu}_\sigma(u_1;Q) \to
  \frac{iQ^0 - D \Q^2}{iQ^0} \,
  {\cal G}^{{\rm R}0}_\sigma(u_1;Q)
\label {eq:GRsub}
\end {equation}
and similarly for ${\cal G}^{{\rm R}\nu}_\tau(u_2;Q')$ in (\ref{eq:Gaarr1}).
Note that the R-charge diffusion constant (\ref{eq:Dvalue})
is $D{=}1$ in our units $2\pi T{=}1$.
As reviewed earlier, our measurements do not need to resolve scales
as small as $1/T$ in order to study jet stopping, and so we may make
small-momentum approximations for the momenta $Q$ and $Q'$ conjugate
to the measurement points $x$ and $y$.  In previous work \cite{adsjet},
we discussed how the corresponding low-momentum
bulk-to-boundary propagators are
\begin {align}
  {\cal G}^{{\rm R}0}_0(\omega,k)
  &\simeq
  \frac{i\omega}{i\omega-k^2}
  - \frac{k^2}{i\omega-k^2} (1-u)^{1-i\omega/2} ,
\\
  {\cal G}^{{\rm R}0}_3(\omega,k)
  &\simeq
  - \frac{ik}{i\omega-k^2}
  + \frac{ik}{i\omega-k^2} (1-u)^{-i\omega/2} .
\end {align}
We also showed that for the purposes of computing
$\langle \Sigma_\Theta(x^0{=}\infty,\x) \rangle$, only near-horizon values of
$u$ were important and that in that limit one could replace%
\footnote{
  See specifically the discussion in sections IV.A.1 and IV.G of
  ref.\ \cite{adsjet}.
}
\begin {subequations}
\label {eq:GRsmall}
\begin {equation}
  {\cal G}^{{\rm R}0}_0(\omega,k)
  \simeq
  \frac{i\omega}{i\omega-k^2}
\end {equation}
and ignore ${\cal G}^{{\rm R}0}_3$ altogether in the calculation:
\begin {equation}
  {\cal G}^{{\rm R}0}_3(\omega,k)
  \simeq
  0.
\end {equation}
\end {subequations}
The same will happen in computing the
correlation of $\Sig$.
Using (\ref{eq:GRsmall}) in (\ref{eq:GRsub}), we see that
replacing the $j$'s by $\Sig$'s in (\ref{eq:basic2})
amounts to replacing
\begin {equation}
  {\cal G}^{{\rm R}\mu}_\sigma(u_1;Q) \to \delta_\sigma^0
  \quad \mbox{and} \quad
  {\cal G}^{{\rm R}\nu}_\tau(u_2;Q') \to \delta_\tau^0
\label {eq:GRsub2}
\end {equation}
in (\ref{eq:Gaarr1}).
Putting everything together and using
\begin {equation}
   \sqrt{-g} \, g^{00} g^{\perp\perp} = \frac{R}{2uf}
\end {equation}
gives
\begin {align}
  \langle \tfrac12 \{ \Sig,\Sig \} \rangle_{\rm(a-1)}
  &\simeq
  \frac{\Aamp^2}{4\gSG^4} \int \frac{du_1}{u_1 f_1} \> \frac{du_2}{u_2 f_2} \>
  \Field^*(x,u_1) \, \tensor\partial_{x^0} \,
  i\Grrperpbulk(x,u_1;y,u_2) \,
  \tensor\partial_{y^0} \, \Field(y,u_2)
\nonumber\\ & \qquad
  + \{x\leftrightarrow y\} ,
\label {eq:basicAT}
\end {align}
where
\begin {equation}
  \Field(x,u)
  \equiv \int_q \calGRup_\perp(u;q) \,
    \tilde\Lambda_L(q-\bar k) \,
    e^{iq\cdot x}
\label {eq:Field}
\end {equation}
describes the bulk vector field created by the source and
\begin {equation}
  \Field^*(x,u)
  = \int_q \calGAup_\perp(u;q) \,
    \tilde\Lambda_L^*(q-\bar k) \,
    e^{-iq\cdot x}
  .
\label {eq:FieldConj}
\end {equation}
For comparison, the corresponding formula for
$\langle \Sig \rangle$ is \cite{adsjet}
\begin {equation}
  \langle\Sig\rangle \simeq
  \frac{{\cal N}_{\rm A}^2}{\gSG^2}
  \int \frac{du}{uf} \>
  \Field^*(x,u) i \tensor\partial_{x^0} \Field(x,u)
  .
\end {equation}

At this point, it is useful to think again in
momentum space, where the operators $\tensor\partial_{x^0}$ and
$\tensor\partial_{y^0}$ in (\ref{eq:basicAT}) become the
$-i(-p-Q_1)_0$ and $-i(Q_2-p)_0$ factors in (\ref{eq:Gaarr1}) respectively.
As mentioned earlier, our resolution requirements permit approximating
$Q$ and $Q'$ as small, in which case these factors may be
approximated by $2i(Q_1)_0$ and $-2i(Q_2)_0$.  Because of the
source factors $\Lambda_L$ in (\ref{eq:basic2}), our calculation only
gets contributions from $(Q_1)_0 \simeq -E$ and $(Q_2)_0 \simeq E$.
The upshot is that the two operators $\tensor\partial_0$ in
(\ref{eq:basicAT}) may each be replaced by $-2iE$:
\begin {equation}
  \langle \tfrac12 \{ \Sig,\Sig \} \rangle_{\rm(a-1)}
  \simeq
  -2E^2
  \frac{\Aamp^2}{\gSG^4} \Re
  \int \frac{du_1}{u_1 f_1} \> \frac{du_2}{u_2 f_2} \>
  \Field^*(x,u_1) \,
  i\Grrperpbulk(x,u_1;y,u_2) \,
  \Field(y,u_2) .
\label {eq:basicAT2}
\end {equation}


\subsection {Factorizing \boldmath$\Grrbulk$}

Let us return to momentum space for $\Grrbulk$:
\begin {multline}
  \langle \tfrac12 \{ \Sig,\Sig \} \rangle_{\rm(a-1)} \simeq
\\
  - 2 E^2
  \frac{\Aamp^2}{\gSG^4} \Re
  \int_p
  \int \frac{du_1}{u_1 f_1} \> \frac{du_2}{u_2 f_2} \>
  \Field^*(x,u_1) \, e^{ip\cdot x} \,
  i\Grrperpbulk(u_1;u_2,-p) \,
  e^{-ip\cdot y} \, \Field(y,u_2) .
\label {eq:basicAT3}
\end {multline}
Using the general formula (\ref{eq:Grrbulk}) for $\Grrbulk$ and
the result (\ref{eq:WRAT}) for $W_{\rm RA}$,
the $u_1$ and $u_2$
integrals factorize as
\begin {equation}
  \langle \tfrac12 \{ \Sig,\Sig \} \rangle_{\rm(a-1)} \simeq
  2E^2 \, \frac{\Aamp^2}{\gSG^2}
  \Re \int_p \frac{\cth(\pi p^0)}{p^0 |b(p)|^2} \,
             {\cal J}^*(p,x) \, {\cal J}(p,y)
  \,,
\label {eq:calJJT}
\end {equation}
where
\begin {align}
  {\cal J}(p,x) &\equiv
  \int \frac{du}{u f} \>
  \left[ \calGRup_\perp(u;p) - \calGAup_\perp(u;p) \right]
  e^{-ip\cdot x} \, \Field(x,u)
\nonumber\\
  &=
  \int_q \tilde\Lambda_L(q-\bar k) \,
  e^{i(q-p)\cdot x}
  \int \frac{du}{u f} \>
   \calGRup_\perp(u;q)
  \left[ \calGRup_\perp(u;p) - \calGAup_\perp(u;p) \right]
  .
\label {eq:Jpx1T}
\end {align}


\subsection {The \boldmath$u$ integral}
\label {sec:uint}

We now evaluate the $u$ integral in (\ref{eq:Jpx1T}),
\begin {equation}
   I_{qp} \equiv
  \int \frac{du}{u f} \>
  \calGRup_\perp(u;q)
  \left[ \calGRup_\perp(u;p) - \calGAup_\perp(u;p) \right] .
\label {eq:IqpdefT}
\end {equation}
Consider the near-horizon (large $\tau$) approximation to
the integrand.  Using (\ref{eq:calGRnear}) for
$\calGRup_\perp$ and using $du/uf \simeq d\tau$,%
\footnote{
   We will consider only real values of $p$ (but complex values of $q$)
   in what follows, and so
   $b(-p)$ and $b^*(p)$ are interchangeable as in (\ref{eq:bconj}).
}
\begin {equation}
  I_{qp} \simeq
  b(q)
  \int_0^\infty d\tau \>
  e^{iq^0\tau}
  \left[ b(p) e^{ip^0\tau} - b^*(p) e^{-ip^0\tau} \right] .
\end {equation}
Recall that $q^0$ is really $q^0+i\epsilon$ because of the
retarded prescription on $q$.  The integral is then
\begin {equation}
  I_{qp} \simeq
  i b(q)
  \left[
     \frac{b(p)}{q^0+p^0+i\epsilon}
     - \frac{b^*(p)}{q^0-p^0+i\epsilon}
  \right] .
\label {eq:IqpT}
\end {equation}

Why is it okay to make the near-horizon approximation for the
bulk-to-bulk propagator?
We are interested in evaluating the correlator (\ref{eq:basicAT3}) of
$\Sig$'s at $x^0,y^0 \to \infty$.  In this limit, ${\cal A}(x,u_1)$
and ${\cal A}(y,u_2)$ are exponentially suppressed everywhere except
very close to the horizon ($\tau \gg 1$) \cite{adsjet}.%
\footnote{
  We can see this in detail from the
  exponential tail derived in our previous work \cite{adsjet} for large
  $X^+ \equiv x^+-\tau(u)$:
  \[
    \Field(x,u)
    \simeq
    i e^{iE [x^-+\tau(u)]}
    \Res\left[{\bar{\cal G}}^{\rm R}_{\perp}(u;q_+^{(1)})\right]
    \Lambda_L^{(2)}\bigl(q_+^{(1)};x^-{+}\tau(u)\bigr) \,
     e^{-\Im(q_+^{(1)}) X^+} .
  \]
  Large $x^0$ for fixed $x^3$ corresponds to large $-x^-$ and $x^+$.
  From the $x^-+\tau$ argument of $\Lambda_L$, we therefore get
  exponential suppression unless $\tau \simeq -x^-$ is also large.
  In this case, $X^+ \simeq 2x^3$.
}

Note from (\ref{eq:calGRnear}) that corrections to the near-horizon limit
formulas for our propagators are suppressed by $e^{-2\tau}$, and so
these corrections vanish in the large-time limit of interest.


\subsection {The \boldmath$q^0$ integral}

Putting the result (\ref{eq:IqpT}) for the $u$ integral back into
eq.\ (\ref{eq:Jpx1T}) for ${\cal J}$ gives
\begin {equation}
  {\cal J}(p,x) =
  i \, b(p)
    \int_q b(q) \, \tilde\Lambda_L(q-\bar k) \,
    \frac{e^{i(q-p)\cdot x}}{q^0+p^0+i\epsilon}
  -
  i \, b^*(p)
    \int_q b(q) \, \tilde\Lambda_L(q-\bar k) \,
    \frac{e^{i(q-p)\cdot x}}{q^0-p^0+i\epsilon}
  \,.
\label {eq:calJ2T}
\end {equation}
Now do the $q^0$ integral by closing the integral in the lower
half-plane,%
\footnote{
   Technically, some care should be taken here because of the behavior of
   $\Lambda_L$ in the complex frequency plane.  Really, one should
   take care to route the closing piece of the
   contour through regions of the complex
   plane where the integrand is exponentially suppressed.
   See Appendix E of ref.\ \cite{adsjet} for related discussion
   (though there the contour is different and serves a
   different purpose).
}
and look at the contribution from picking
up the explicit poles above:
\begin {multline}
  {\cal J}(p,x) \simeq
  e^{2ip^0 x^0} b(p)
    \int_\q b(-p^0,\q) \, \tilde\Lambda_L(-p^0-E,\q-\bar\k) \,
    e^{i(\q-\p)\cdot\x}
\\
  -
  b^*(p)
    \int_\q b(p^0,\q) \, \tilde\Lambda_L(p^0-E,\q-\bar\k) \,
    e^{i(\q-\p)\cdot\x}
  \,.
\label {eq:Jsimp0}
\end {multline}
Before we use this expression, we should consider what other
singularities contribute to the integral.  In particular,
$b(q)$ has poles in the lower-half $q^0$ plane, corresponding to
quasi-normal modes.  When evaluated at these poles, the
$e^{-iq^0 x^0}$ term in the integrand will give a suppression
factor of order
\begin {equation}
   e^{x^0 \Im q_{\rm pole}^0} .
\end {equation}
(Keep in mind that $\Im q_{\rm pole}^0$ is negative.)
But we are interested specifically
in the $x^0 \to \infty$ limit for evaluating $\Sig$,
and so these contributions will vanish.  Because $\tilde\Lambda_L$
only has support for $q_- \simeq E$, where the poles in $q_+$
have imaginary parts of $O(E^{-1/3})$ \cite{adsjet}, the relevant
poles in $q^0 = q_- - q_+$ have imaginary parts of $-O(E^{-1/3})$.

Because of the support of the envelope factor $\tilde\Lambda_L$ in
the first term of (\ref{eq:Jsimp0}), the $e^{2ip^0 x^0}$ phase factor
in that term has $p^0 \simeq E$ and so is highly oscillatory.
Remember that we are only trying to resolve distance and
time scales of the observables on scales large compared to $1/T$.
If we smear out our observables $\hat\Sigma_\Theta(x)$ and
$\hat\Sigma_\Theta(y)$ over such scales,
then the contribution of such a highly-oscillating phase will be
smeared away.%
\footnote{
  More specifically, suppose we replace $\hat\Sigma_\Theta(x)$ by
  \[
   [\hat\Sigma_\Theta(x)]_{\rm smeared}
   \equiv
   \int d^4(\Delta x) \,
   \hat\Sigma_\Theta(x+\Delta x) \,
   \frac{
     e^{-(\Delta x^0)^2/\ell_{\rm smear}^2}
     e^{-|\Delta\x|^2/\ell_{\rm smear}^2}
   }{\pi \ell_{\rm smear}^2}
 \]
 as discussed in sec.\ II B of ref.\ \cite{adsjet}, where the smearing
 distance $\ell_{\rm smear}$ is chosen large compared to microscopic
 scales such as $1/E$ and $1/T$ but small compared to the scales we're
 interested in resolving, such as stopping distances.
 If one applies this procedure to something that behaves like
 $\exp(i2E x^0)$, then one obtains an exponentially small result.
}
As a result, we may drop the first term of (\ref{eq:Jsimp0}) and
simply write
\begin {equation}
  {\cal J}(p,x) \simeq
  -
  b^*(p)
    \int_\q b(p^0,\q) \, \tilde\Lambda_L(p^0-E,\q-\bar\k) \,
    e^{i(\q-\p)\cdot\x}
  \,.
\label {eq:Jsimp}
\end {equation}


\subsection {Assembling the pieces}

Now use the expression (\ref{eq:Jsimp}) for ${\cal J}$ in
the correlator (\ref{eq:calJJT}) of $\Sig$'s:
\begin {align}
  \langle \tfrac12 \{ \Sig,\Sig \} \rangle_{\rm(a-1)} & \simeq
  2E^2 \, \frac{\Aamp^2}{\gSG^2}
  \Re \int_p \frac{\cth(\pi p^0)}{p^0} e^{i\p\cdot(\x-\y)}
       \int_{\q\q'} e^{-i\q\cdot\x} e^{-i\q'\cdot\y}
\nonumber\\ & \qquad \times
        b^*(p^0,\q) \, b(p^0,\q') \,
        \tilde\Lambda_L^*(p^0{-}E,\q{-}\bar\k) \,
        \tilde\Lambda_L(p^0{-}E,\q'{-}\bar\k)
  .
\end {align}
Now note that the $\p$ integral just gives $\delta^{(3)}(\x-\y)$:
\begin {align}
  \langle \tfrac12 \{ \Sig,\Sig \} \rangle_{\rm(a-1)} & \simeq
  2E^2 \, \frac{\Aamp^2}{\gSG^2}
  \, \delta^{(3)}(\x-\y)
  \Re \int_{p^0} \frac{\cth(\pi p^0)}{p^0}
       \int_{\q\q'} e^{-i\q\cdot\x} e^{-i\q'\cdot\y}
\nonumber\\ & \qquad \times
        b^*(p^0,\q) \, b(p^0,\q') \,
        \tilde\Lambda_L^*(p^0{-}E,\q{-}\bar\k) \,
        \tilde\Lambda_L(p^0{-}E,\q'{-}\bar\k)
  .
\label {eq:final}
\end {align}
The right-hand side vanishes for $\x \not= \y$.  Recall that
throughout we have made approximations that blur our resolution
of $\x$ and $\y$ on the scale $1/T$.

And so we have answered the question we set out to answer:
For $|\x-\y|$ large compared to $1/T$,
the correlation is exponentially small (coming from
all the exponentially-small corrections that we dropped throughout).


\begin{acknowledgments}

We gratefully acknowledge Chaolun Wu for his role in deriving the
Heun function expression in footnote \ref{foot:heun2}.
We thank Sangyong Jeon and Derek Teaney for useful conservations.
This work was supported, in part, by the U.S. Department
of Energy under Grant No.~DE-FG02-97ER41027.

\end{acknowledgments}

\appendix

\section {Generalization to jets with large R charge}
\label{app:Rcharge}

In this appendix, we discuss how the argument in the main text is
modified if one uses a source operator with R charge larger than
one.  For simplicity, we will consider scalar operators.
An example is $\tr(X^\Delta)$, where $X$ is any one of the three
complex scalar fields in ${\cal N}{=}4$ SYM.  This operator has
conformal dimension $\Delta$.  The R charge $J$ of $\tr(X^\Delta)$
is $\Delta$ under
a U(1) subgroup of the SU(4) R charge symmetry, which is the
subgroup that we will choose for our measurement operators
$j^0(x)$ and $j^0(y)$.  [That is, it is the U(1) subgroup that
corresponds to $\tau^3/2$ in the main text.]

In what follows, we more generally consider taking our source operator
to be any
scalar BPS operator of conformal dimension $\Delta$ and
R charge $J$ under a U(1) R symmetry subgroup. In the
5-dimensional bulk these correspond to scalar fluctuations of mass
$m$ such that
\begin {equation}
   \Delta=\tfrac{d}{2}+\sqrt{(\tfrac{d}{2})^2+m^2 R^2} ,
\end {equation}
where $d{=}4$ is the number of boundary space-time dimensions.
The important property of the bulk-to-boundary scalar propagator is that
it takes the form
\begin {equation}
  \calGR(u,q)
   = b(q) \, (1-u)^{-i\omega/2} [1+O(1-u)]
   = b(q) \, e^{i\omega\tau} [1+O(e^{-2\tau})]
\end {equation}
near the horizon, similar to (\ref{eq:calGRnearu}) and
(\ref{eq:calGRnear}) but with a different function
$b(q)$ than in the transverse gauge boson case.
Near the boundary $u{=}0$ it behaves like
\begin {equation}
  \calGR(u,q) \propto u^{(d-\Delta)/2} [1 + O(u)] .
\end {equation}
The divergence of the bulk-to-boundary
propagator at the horizon for $d>\Delta$ (related to the need to
renormalize such operators) will not have any effect on our calculation,
which will be controlled by the near-horizon behavior.%
\footnote{
   \label {foot:heun2}
   The full expression for the scalar bulk-to-boundary propagator
   can be written as
   \[
      \calGR(u,q) = (4u)^{\Delta_-/2}(1-u)^{-i\omega/2}(1+u)^{\omega/2}
                  \frac{h(u)}{h(0)} ,
   \]
   where
   \[
     h(u) =
    \Hl(
       2,
       q^2 + \tfrac{\Delta_-^2}4 - i\Delta_-\omega + \tfrac{(1+i)\omega}2
           - \tfrac{i\omega^2}2 ;
       \tfrac{\Delta_-}2 + \tfrac{(1-i)\omega}2 ,
       \tfrac{\Delta_-}2 + \tfrac{(1-i)\omega}2 ,
       1-i\omega,
       \Delta_--1;
       1-u) ,
   \]
   $\Delta_- = \tfrac{d}{2}-\sqrt{(\tfrac{d}{2})^2+m^2 R^2} = d-\Delta$,
   and $\Hl$ is the Heun function \cite{Heun}.
   (Compare to footnote \ref{foot:heun}.)
   Here we have naively
   normalized the bulk-to-boundary propagator so that the coefficient
   of the small-$u$ behavior
   $(4u)^{\Delta_-/2}=(4u)^{(d-\Delta)/2} = z^{d-\Delta}$
   is precisely 1 instead of some
   more complicated normalization involving regularization of the
   boundary to some small, finite value
   $z_{\rm B}$ of $z \equiv \sqrt{4u}$.
   As argued by ref.\ \cite{Freedman},
   the naive normalization of the bulk-to-boundary
   propagator can be used
   when computing $n$-point functions with $n>2$.
}

In the vicinity of the horizon,
the bulk-to-bulk retarded and symmetrized propagators take the form
\begin{align}
  i \GRbulk(\tau_1;\tau_2,q)
  &= 2 \, \frac{\gSG^2}{\omega}e^{i\omega\tau_>}
     [-e^{i\delta(q)}e^{i\omega\tau_<}+e^{-i\omega\tau_<}]
\\
  i\Grrbulk(\tau_1;\tau_2,q)
  &= 4 \gSG^2 \, \frac{\cth(\pi\omega)}{\omega}
     \left[
       - \cos\Bigl(\omega(\tau_2+\tau_1) + \delta(q)\Bigr)
       + \cos\Bigl(\omega(\tau_2-\tau_1)\Bigr)
     \right] ,
\end{align}
where, as before,
$e^{i\delta(q)} \equiv b(q)/b(-q)$. These formulas differ from the
transverse boson case (\ref{eq:GRnear}) and (\ref{eq:Grrnear})
by a factor of 2 normalization
(and $\delta(q)$ is a different function of $q$).

For a scalar field with R charge $J$, the same real-time Witten diagram
(a-1) as in the main text gives
\begin {align}
   \left[
    (G_{\rm aarr})^{(33)\mu\nu}
   \right]_{\rm(a-1)}
   &=
   -i J^2
\nonumber\\ & \quad \times
   \frac{2i(-p-Q_1)_\rho}{\gSG^2 R^3}
   \int du_1 \> (\sqrt{-g} g^{\rho\sigma})_1 \,
   \calGA(Q_1,u_1) \, {\cal G}_\sigma^{{\rm R}\mu}(Q,u_1)
\nonumber\\ & \quad \times
   \frac{2i(Q_2-p)_\xi}{\gSG^2 R^3}
   \int du_2 \> (\sqrt{-g} g^{\xi\tau})_2 \,
   \calGA(Q_2,u_2) \, {\cal G}_\tau^{{\rm R}\nu}(Q',u_2)
\nonumber\\ & \quad \times
   i \Grrbulk(u_1;u_2,-p) \biggl|_{p=Q_1+Q=-(Q_2+Q')}
\nonumber\\ &
   + \{Q,\mu\leftrightarrow Q',\nu\} .
\end {align}
Substituting the metric factors and the various propagators leads to
\begin {align}
  \langle \tfrac12 \{ \Sig,\Sig \} \rangle_{\rm(a-1)}
  &\simeq
  \frac{\Aamp^2 J^2}{32\gSG^4}
  \int \frac{du_1}{u_1^2 f_1} \> \frac{du_2}{u_2^2 f_2} \>
  \Field^*(x,u_1) \, \tensor\partial_{x^0} \,
  i\Grrbulk(x,u_1;y,u_2) \,
  \tensor\partial_{y^0} \, \Field(y,u_2)
\nonumber\\ & \qquad
  + \{x\leftrightarrow y\} ,
\end {align}
where now
\begin{equation}
\Field(x,u) = \int_q \calGR(u;q) \, \Lambda_L(q-\bar k) \, e^{i q\cdot x}.
\end{equation}
The same considerations we have previously made apply: namely we can
approximate the action of $\tensor\partial_{x^0}$ and $\tensor\partial_{y^0}$
by $-2iE$ factors, substitute the factorized form of the symmetrized
propagator, and evaluate the subsequent $u$ integrals,
\begin{equation}
  I_{qp} = \int \frac{du}{u^2 f} \>
           \calGR(u;q) \bigl[\calGR(u;p)-\calGA(u;p)\bigr].
\end{equation}
$I_{qp}$ above
differs from (\ref{eq:IqpdefT}) by a factor of $1/u$, but in the
near-horizon approximation $u\simeq 1$ and $du/(u^2 f)\simeq d\tau$ it
becomes formally identical to (\ref{eq:IqpT}):
\begin{equation}
  I_{qp} \simeq
  i b(q)
  \bigg[\frac{b(p)}{q^0+p^0+i\epsilon}-\frac{b^*(p)}{q^0-p^0+i\epsilon}\bigg].
\end{equation}
The final result is then the same as (\ref{eq:final}) but with the
substitution $\Aamp^2 \to \Aamp^2 J^2/4$ in the overall normalization.
The conclusion is the same as for the case analyzed in the main text:
for late times
($x^0, y^0\to\infty)$ the integrated charge deposition correlator is
exponentially suppressed for $|\bm x - \bm y| \gg 1/T$.


\section {Relation of our near-horizon \boldmath$\Grrbulk$ to
          ref.\ \cite{CCT}}
\label {app:CCT}

Caron-Huot, Chesler, and Teaney \cite{CCT} studied correlators of
fluctuations of the position $\hat x(t,r)$ of a classical string,
where $r$ was their coordinate for the 5th dimension, running from
$r{=}1$ at the horizon to $r{=}\infty$ at the boundary.  One of their
results was
\begin {equation}
  G_{\rm rr}(t_1,r_1;t_2,r_2) =
  \int dt_1'\>dt_2' \>
  \bigl[-G_{\rm ra}(t_1,r_1;t_1',r_{\rm h}) \bigl]
  \bigl[-G_{\rm ra}(t_2,r_2;t_2',r_{\rm h}) \bigl]
  G_{\rm rr}^{\rm h}(t_1',t_2')
\label {eq:GrrCCT}
\end {equation}
for $r_1,r_2 > r_{\rm h}$, where $r_{\rm h} = 1 + \epsilon$ with
$\epsilon$ very small defined a ``stretched'' horizon and where
$G_{\rm rr}^{\rm h}(t_1',t_2')$ was a type of correlator on that
stretched horizon.  For our purposes here, we may treat
$\epsilon$ as infinitesimal.  They found that
\begin {equation}
   G_{\rm rr}^{\rm h} =
   - \frac{\eta}{\pi} \partial_{v_1} \partial_{v_2}
     \ln|1-e^{-2\pi T(v_1-v_2)}| ,
\label {eq:GrrhCCT}
\end {equation}
where $v$ is Eddington-Finkelstein time
\begin {equation}
   v \equiv t + \frac{1}{\pi T} \int \frac{dr}{f r^2}
\end {equation}
evaluated in (\ref{eq:GrrhCCT}) at $r_1=r_2=r_{\rm h}$.
So (\ref{eq:GrrhCCT}) can be rewritten
as
\begin {equation}
   G_{\rm rr}^{\rm h} =
   - \frac{\eta}{\pi} \partial_{t_1} \partial_{t_2}
     \ln|1-e^{-2\pi T(t_1-t_2)}| .
\label {eq:Grrh}
\end {equation}
Their definition of $G_{\rm rr}$ as $\tfrac12 \langle\{ O(1),O(2)\}\rangle$
for an operator $O$ is $i/2$ times ours, but we will not
worry about overall normalization factors in this discussion
since the final formulas for $G_{\rm rr}$ also depend
how one normalizes the operators of interest
to a particular problem.

We will now see that these same equations describe formulas in our
paper (up to overall normalization)
provided we generalize (\ref{eq:GrrhCCT}) to fields with
$\x$ dependence,
\begin {equation}
  \Grrbulk(x_1,r_1;x_2,r_2) =
  \int dx_1'\>dx_2' \>
  \bigl[-\Gbulk_{\rm ra}(x_1,r_1;x_1',r_{\rm h}) \bigl]
  \bigl[-\Gbulk_{\rm ra}(x_2,r_2;x_2',r_{\rm h}) \bigl]
  G_{\rm rr}^{\rm h}(x_1',x_2'),
\label {eq:GrrCCT2}
\end {equation}
and keep (\ref{eq:Grrh}) the same.

Fourier transforming (\ref{eq:Grrh}) gives
\begin {equation}
   G_{\rm rr}^{\rm h}(q) \propto \omega \cth(\tfrac12 \beta \omega)
   .
\label {eq:Grrhw}
\end {equation}
The Fourier transform of (\ref{eq:GrrCCT2}) is
\begin {align}
  \Grrbulk(r_1;r_2,q)
  &=
  \Gbulk_{\rm ra}(r_1,-q;r_{\rm h}) \,
  \Gbulk_{\rm ra}(r_2,q;r_{\rm h}) \,
  G_{\rm rr}^{\rm h}(q)
\nonumber\\
  &=
  [\GRbulk(r_{\rm h};r_1,q)]^* \,
  \GRbulk(r_{\rm h};r_2,q)     \,
  G_{\rm rr}^{\rm h}(q)
  ,
\label {eq:GrrCCT3}
\end {align}
given our convention in this paper that $G_{\rm R}$ means $G_{\rm ar}$.

We can now proceed generally or specialize to the case where
$r_1$ and $r_2$ are near the horizon (though not as near as $r_{\rm h}$).
In the latter case, simply plugging in our near-horizon formula
(\ref{eq:GRnear}) for $\GRbulk$ and (\ref{eq:Grrhw}) into
(\ref{eq:GrrCCT3}) gives
\begin {equation}
  \Grrbulk
  \propto 
  \frac{\cth(\tfrac12 \beta \omega)}{\omega}
  \bigl\{e^{i\omega\tau_{\rm h}}
    [-e^{i\delta(q)} e^{i\omega \tau_1} + e^{-i\omega\tau_1}]
  \bigr\}^*
  \bigl\{e^{i\omega\tau_{\rm h}}
    [-e^{i\delta(q)} e^{i\omega \tau_2} + e^{-i\omega\tau_2}]
  \bigr\} ,
\end {equation}
which reproduces our near-horizon result (\ref{eq:Grrnear}) for
$\Grrbulk$.  In the more general case, use (\ref{eq:GRbulk}) instead
of (\ref{eq:GRnear}) to obtain
\begin {align}
  \Grrbulk
  &\propto 
  \omega \cth(\tfrac12 \beta \omega)
  \Bigl\{
    \frac{\calGR(u_{\rm h};q)}{W_{\rm RA}(q)}
    \bigl[ \calGR(u_1;q) - \calGA(u_1;q) \bigr]
  \Bigr\}^*
  \Bigl\{
    \frac{\calGR(u_{\rm h};q)}{W_{\rm RA}(q)}
    \bigl[ \calGR(u_2;q) - \calGA(u_2;q) \bigr]
  \Bigr\}
\nonumber\\
  &=
  \frac{-\omega|\calGR(u_{\rm h};q)|^2}{W_{\rm RA}^*(q)}
  \frac{\cth(\tfrac12 \beta \omega)}{W_{\rm RA}(q)}
  \bigl[ \calGR(u_1;q) - \calGA(u_1;q) \bigr]
  \bigl[ \calGR(u_2;q) - \calGA(u_2;q) \bigr]
  .
\end {align}
This reproduces (\ref{eq:Grrbulk}) with the aid of
(\ref{eq:calGRnearu}) for $\calGR(u_{\rm h};q)$ and
(\ref{eq:WRAT}).


\section {\boldmath$T{=}0$ analysis of
          \boldmath$\tfrac12\langle \{j^\mu,j^\nu\}\rangle$}
\label {app:Tzero}

It's useful to see how the formalism used in this paper works in the
zero-temperature case.  Among other things, it provides a useful double
check of the normalization of some of our basic equations.

There is no thermalization and ``charge deposition'' in vacuum, and
so we will study
$\tfrac12 \langle \{ j^\mu(x), j^\nu(y) \}\rangle$
instead of
$\tfrac12 \langle \{ \hat\Sigma_\Theta(x),
                     \hat\Sigma_\Theta(y) \}\rangle$.
At zero temperature, a jet excitation simply propagates forever.
The excitation's total current should not change with time,
and so it is natural to expect that
\begin {align}
   \Delta \Bigl\langle \tfrac12 \bigl\{ \int_{\x} j^\mu(x^0,\x),
        \int_{\y} j^\nu(y^0,\y) \bigr\}\Bigr\rangle_{\rm jet}
   &\simeq \Bigl\langle \int_{\x} j^\mu(x^0,\x) \Bigr\rangle_{\rm jet}
           \Bigl\langle \int_{\y} j^\nu(y^0,\y) \Bigr\rangle_{\rm jet}
\nonumber\\
   &\simeq e_{\rm jet}^2 \frac{\bar k^\mu \bar k^\nu}{E^2} \,
      \theta(x^0) \, \theta(y^0)
\label {eq:checkT0}
\end {align}
at zero temperature.
The goal of this appendix is to show that the methods used in this paper
indeed reproduce this expected result.
Here, $e_{\rm jet}=1$ as in the main text.
Above, the $\Delta$ indicates we subtract (vacuum)
fluctuations as in (\ref{eq:fd}), analogous to (\ref{eq:jjsubtract}).

In the following, we will write the zero-temperature metric as
\begin {equation}
  ds^2
        =  \frac{R^2}{4} \left[ \frac{1}{\bar u}(-dt^2 + d\x^2)
         + \frac{1}{\bar u^2} \, d\bar u^2 \right] ,
\end {equation}
where $\bar u$ corresponds to our earlier $u/(2\pi T)^2$ and runs from
zero to infinity.

Start from (\ref{eq:basic2}) and (\ref{eq:Gaarr1}).  As discussed
in ref.\ \cite{adsjet}, the low-wavenumber approximation for the
observable corresponds to
\begin {equation}
   {\cal G}^{\rm R\mu}_\sigma(\bar u;Q) \simeq \delta^{\mu}_{\sigma}
\label {eq:calGapproxT0}
\end {equation}
at zero temperature.
In this approximation, (\ref{eq:calfluctdisp}) gives
\begin {equation}
   {\cal G}^{\rm rr\mu}_\sigma(\bar u;Q) \simeq 0 ,
\end {equation}
and so we can ignore every diagram in fig.\ \ref{fig:diagsra} except
(a-1).
Combining (\ref{eq:basic2}) and (\ref{eq:Gaarr1})
using (\ref{eq:calGapproxT0}) gives
\begin {align}
  \Delta \langle \tfrac12 \{ j^\mu ,j^\nu \} \rangle
  &\simeq
  \frac{\Aamp^2}{4\gSG^4}
  \int \frac{d\bar u_1}{\bar u_1} \> \frac{d\bar u_2}{\bar u_2} \>
  \Field^*(x,\bar u_1) \, \tensor\partial_{x^\mu} \,
  i\Grrperpbulk(x,\bar u_1;y,\bar u_2) \,
  \tensor\partial_{y^\nu} \, \Field(y,\bar u_2)
\nonumber\\ & \qquad
  + \{x\leftrightarrow y\} .
\end {align}
Note that the zero-temperature formula
for $\tfrac12 \langle \{ j^0, j^0 \}\rangle$
happens to have exactly the same form as the
finite-temperature formula
(\ref{eq:basicAT}) for
$\tfrac12 \langle \{ \hat\Sigma_\Theta, \hat\Sigma_\Theta \}\rangle$
(noting that $f{\to}1$ at zero temperature).
Following the same line of argument that led to (\ref{eq:basicAT3})
then gives the analogous zero-temperature formula
\begin {multline}
  \Delta \langle \tfrac12 \{ j^\mu,j^\nu \} \rangle \simeq
\\
  - 2 \bar k^\mu \bar k^\nu
  \frac{\Aamp^2}{\gSG^4} \Re
  \int_p
  \int \frac{d\bar u_1}{\bar u_1} \> \frac{d\bar u_2}{\bar u_2} \>
  \Field^*(x,\bar u_1) \, e^{ip\cdot x} \,
  i\Grrperpbulk(\bar u_1;\bar u_2,-p) \,
  e^{-ip\cdot y} \, \Field(y,\bar u_2) .
\end {multline}
It's helpful at this point to have the explicit formulas for
the transverse-polarized bulk-to-boundary propagators in
4-momentum space, which are
\begin {equation}
   {\cal G}(\bar u;q) = \sqrt{4 \bar u q^2} \, K_1(\sqrt{4 \bar u q^2}) ,
\label {eq:calGT0}
\end {equation}
where $K_n$ is the modified Bessel function of the second kind.
$q^0$ in this formula means $q^0 + i\epsilon$ for the retarded
propagator and $q^0-i\epsilon$ for the advanced propagator.
The Wronskian (\ref{eq:WRA}), which is most easily evaluated
in the $\bar u \to 0$ limit, is then
\begin {equation}
   W_{\rm RA}(q) = i \pi R q^2 \operatorname{sign}(q^0) .
\label {eq:WRAT0}
\end {equation}
Note also that the zero-temperature ($\beta\to\infty$)
limit of the
fluctuation-dissipation relation (\ref{eq:fluctdisp})
is
\begin {equation}
   i G_{rr}(q) =
   \operatorname{sign}(q^0) \left[
      i G_{\rm R}(q) - i G_{\rm A}(q)
   \right] .
\end {equation}
Comparing (\ref{eq:WRAT0}) to the finite-temperature formula
(\ref{eq:WRAT}), the upshot is that the analysis of
$\langle\tfrac12 \{ j^\mu, j^\nu \}\rangle$ at
zero temperature goes through just as the finite-temperature
derivation of (\ref{eq:calJJT}) for
$\langle\tfrac12 \{ \Sig^\mu, \Sig^\nu \}\rangle$ but
with the replacement
\begin {equation}
   \frac{\cth(\tfrac12 \beta p^0)}{p^0 |b(p)|^2}
   \to
   - \frac{1}{\pi p^2} \,,
\end {equation}
to give
\begin {equation}
  \Delta \langle \tfrac12 \{ j^\mu,j^\nu \} \rangle \simeq
  - 2\bar k^\mu \bar k^\nu \, \frac{\Aamp^2}{\gSG^2}
  \Re \int_p \frac{1}{\pi p^2} \,
             {\cal J}^*(p,x) \, {\cal J}(p,y)
  \,,
\label {eq:basicJJT0}
\end {equation}
with ${\cal J}$ defined as before in (\ref{eq:Jpx1T}).
Using (\ref{eq:calGT0}),
the desired $u$-integral (\ref{eq:IqpdefT}) of ${\cal J}$ is given
explicitly by
\begin {align}
   I_{qp} &=
    4 \pi i \, \theta(-p^2) \, \sqrt{-p^2 q^2}
    \int_0^\infty d\bar u \>
    K_1(\sqrt{4 \bar u q^2}) \, J_1(\sqrt{-4 \bar u p^2})
\nonumber\\
   &=
    2\pi i \, \frac{p^2}{p^2-q^2} \, \theta(-p^2)
\label {eq:Iqp}
\end {align}
with $q^0$ interpreted as $q^0 + i\epsilon$.
Substitution into (\ref{eq:Jpx1T}) gives
\begin {equation}
  {\cal J}(p,x) =
  -2\pi i p^2 \, \theta(-p^2) \int_q \tilde\Lambda_L(q-\bar k) \,
  \frac{e^{i(q-p)\cdot x}}{q^2-p^2}
\label {eq:calJ3T0}
\end {equation}
as the zero-temperature analog of (\ref{eq:calJ2T}),
but note that we have not made anything analogous to the ``near-horizon''
approximation in the zero-temperature analysis.

Since our sanity check (\ref{eq:checkT0})
of results and normalizations in this appendix involves integrating
over 3-position, it's convenient at this point to integrate
(\ref{eq:calJ3T0}) over $\x$ to get
\begin {equation}
  \int_\x {\cal J}
  = i p^2 \theta(-p^2)
    \int dq^0 \>
    \frac{e^{-i(q^0-p^0)x^0}}{(q^0+i\epsilon)^2-(p^0)^2} \,
    \tilde\Lambda_L(q^0{-}E,\p^\perp,p^3{-}E) .
\label {eq:calJ4T0}
\end {equation}
Following similar approximations as from (\ref{eq:calJ2T}) to
(\ref{eq:Jsimp}), pick up only the explicit poles in
(\ref{eq:calJ4T0}) and then throw away highly-oscillatory terms
in the result.  This yields
\begin {align}
  \int_\x {\cal J}
  &\simeq \frac{\pi p^2}{p^0} \, \theta(-p^2) \,
    \theta(x^0) \,
    \tilde\Lambda_L(p-\bar k)
\nonumber\\
  &\simeq 4\pi p_+ \, \theta(-p_+) \,
    \theta(x^0) \,
    \tilde\Lambda_L(p-\bar k) .
\end {align}
Using this expression in (\ref{eq:basicJJT0}) gives
\begin {align}
  \Delta\langle \tfrac12
     \{ \smallint\nolimits_\x j^\mu,\smallint\nolimits_\y j^\nu\}
  \rangle
  &\simeq
  - 32 \pi\bar k^\mu \bar k^\nu \, \theta(x^0) \, \theta(y^0) \,
  \frac{\Aamp^2}{\gSG^2}
  \int_p \theta(-p_+) \, \frac{(p_+)^2}{p^2}
  \left| \tilde\Lambda_L(p-\bar k) \right|^2
\nonumber\\
  &\simeq
  {\cal Q}
  \frac{\bar k^\mu \bar k^\nu}{E^2} \, \theta(x^0) \, \theta(y^0) ,
\label {eq:T0final}
\end {align}
where
\begin {equation}
   {\cal Q} \simeq
   8 \pi E \, \frac{\Aamp^2}{\gSG^2}
   \int_q \>\theta(-q_+) \,
   |q_+| \, \bigl|\tilde\Lambda_L(q)\bigr|^2 .
\end {equation}
This formula for ${\cal Q}$ is equivalent to that found in
ref.\ \cite{adsjet} for the average charge created by our source
operator.  Dividing both sides of (\ref{eq:T0final}) by
${\cal Q}$ and invoking (\ref{eq:anglejet}) finally gives us
(\ref{eq:checkT0}), as expected.


{}


\begin{thebibliography}{}

\bibitem{SonStarinets}
  D.~T.~Son and A.~O.~Starinets,
  ``Viscosity, Black Holes, and Quantum Field Theory,''
  Ann.\ Rev.\ Nucl.\ Part.\ Sci.\  {\bf 57}, 95 (2007)
  [arXiv:0704.0240].

\bibitem{BRSSS}
  R.~Baier, P.~Romatschke, D.~T.~Son, A.~O.~Starinets and M.~A.~Stephanov,
  ``Relativistic viscous hydrodynamics, conformal invariance, and holography,''
  JHEP {\bf 0804}, 100 (2008)
  [arXiv:0712.2451].

\bibitem{adsjet}
  P.~Arnold, D.~Vaman,
  ``Jet quenching in hot strongly coupled gauge theories revisited:
  3-point correlators with gauge-gravity duality,''
  JHEP {\bf 1010}, 099 (2010)
  [arXiv:1008.4023].

\bibitem{adsjet2}
  P.~Arnold, D.~Vaman,
  ``Jet quenching in hot strongly coupled gauge theories simplified,''
  JHEP {\bf 1104}, 027 (2011) 
  [arXiv:1101.2689].

\bibitem{qm11}
  P.~Arnold, D.~Vaman,
  ``Some new results for `jet' stopping in AdS/CFT,''
  arXiv:1106.1680,
  an abridged version to appear in J. Phys.\ G.

\bibitem{AVWXhydro2}
  P.~Arnold, D.~Vaman, C.~Wu, W.~Xiao,
  ``Second order hydrodynamic coefficients from
    3-point stress tensor correlators via AdS/CFT,''
  JHEP {\bf 1110}, 033 (2011)
  [arXiv:1105.4645].

\bibitem{SShydro2}
  O.~Saremi, K.~A.~Sohrabi,
  ``Causal three-point functions and nonlinear second-order
    hydrodynamic coefficients in AdS/CFT,''
  arXiv:1105.4870.

\bibitem{HattaUeda}
  Y.~Hatta, T.~Ueda,
  ``Soft photon anomaly and gauge/string duality,''
  Nucl.\ Phys.\  {\bf B837}, 22-39 (2010)
  [arXiv:1002.3452].

\bibitem{SquarePenrose}
  L.~Fidkowski, V.~Hubeny, M.~Kleban and S.~Shenker,
  ``The black hole singularity in AdS/CFT,''
  JHEP {\bf 0402}, 014 (2004)
  [arXiv:hep-th/0306170].

\bibitem{CCT}
  S.~Caron-Huot, P.~M.~Chesler, D.~Teaney,
  ``Fluctuation, dissipation, and thermalization in non-equilibrium
    AdS$_5$ black hole geometries,''
  [arXiv:1102.1073].

\bibitem{CJK}
  P.~M.~Chesler, K.~Jensen and A.~Karch,
  ``Jets in strongly-coupled ${\cal N}{=}4$ super Yang-Mills theory,''
  Phys.\ Rev.\  D {\bf 79}, 025021 (2009)
  [arXiv:0804.3110].

\bibitem{Rdiffusion}
  G.~Policastro, D.~T.~Son and A.~O.~Starinets,
  ``From AdS/CFT correspondence to hydrodynamics,''
  JHEP {\bf 0209}, 043 (2002)
  [arXiv:hep-th/0205052].


\bibitem{Kubo}
  R. Kubo, ``The fluctuation-dissipation theorem,''
  Rep.\ Prog.\ Phys.\ {\bf 29}, 244 (1966).

\bibitem{WangHeinz}
  E.~Wang and U.~W.~Heinz,
  ``A generalized fluctuation-dissipation theorem for nonlinear response
  functions,''
  Phys.\ Rev.\  D {\bf 66}, 025008 (2002)
  [arXiv:hep-th/9809016].

\bibitem{SonTeaney}
  D.~T.~Son, D.~Teaney,
  ``Thermal Noise and Stochastic Strings in AdS/CFT,''
  JHEP {\bf 0907}, 021 (2009)
  [arXiv:0901.2338].

\bibitem{HydroTails}
  S.~Caron-Huot, O.~Saremi,
  ``Hydrodynamic Long-Time tails From Anti de Sitter Space,''
  JHEP {\bf 1011}, 013 (2010)
  [arXiv:0909.4525].

\bibitem{GibbonsPerry}
 G.~W.~Gibbons, M.~J.~Perry,
  Phys.\ Rev.\ Lett.\  {\bf 36}, 985 (1976).

\bibitem{HerzogSon}
  C.~P.~Herzog and D.~T.~Son,
  ``Schwinger-Keldysh propagators from AdS/CFT correspondence,''
  JHEP {\bf 0303}, 046 (2003)
  [arXiv:hep-th/0212072].

\bibitem{Kamenev}
   A. Kamenev and A. Levchenko,
   ``Keldysh technique and non-linear $\sigma$-model:
   basic principles and applications,''
   Adv.\ in Phys.\ {\bf 58} (2009) 197.
   [arXiv:0901.3586].

\bibitem{ChaikinLubensky}
  P.~M.~Chaikin and T.~C.~Lubensky,
  {\it Principles of Condensed Matter Physics}
  (Cambridge University Press, 1995).

\bibitem {Veltman}
  M.~J.~G.~Veltman,
  ``Unitarity and causality in a renormalizable field theory with unstable
  particles,''
  Physica {\bf 29}, 186 (1963);
  G.~'t Hooft and M.~J.~G.~Veltman,
  ``Diagrammar,''
  NATO Adv.\ Study Inst.\ Ser.\ B Phys.\  {\bf 4}, 177 (1974).

\bibitem{vertical}
  F.~Gelis,
  ``A New approach for the vertical part of the contour in
    thermal field theories,''
  Phys.\ Lett.\  {\bf B455}, 205-212 (1999)
  [hep-ph/9901263];
  ``The Effect of the vertical part of the path on the real time
    Feynman rules in finite temperature field theory,''
  Z.\ Phys.\  {\bf C70}, 321-331 (1996)
  [hep-ph/9412347].

\bibitem{3point}
  E.~Barnes, D.~Vaman, C.~Wu, P.~Arnold,
  ``Real-time finite-temperature correlators from AdS/CFT,''
  Phys.\ Rev.\  {\bf D82}, 025019 (2010).
  [arXiv:1004.1179].

\bibitem{SvR1}
  K.~Skenderis and B.~C.~van Rees,
  ``Real-time gauge/gravity duality,''
  Phys.\ Rev.\ Lett.\  {\bf 101}, 081601 (2008)
  [arXiv:0805.0150].

\bibitem{SvR2}
  K.~Skenderis and B.~C.~van Rees,
  ``Real-time gauge/gravity duality: Prescription, Renormalization and
  Examples,''
  JHEP {\bf 0905}, 085 (2009)
  [arXiv:0812.2909].

\bibitem {Romans}
  L.~J.~Romans,
  ``Gauged $N{=}4$ Supergravities In
    Five-Dimensions And Their Magnetovac Backgrounds,''
  Nucl.\ Phys.\  B {\bf 267}, 433 (1986).

\bibitem{SU2U1}
  H.~Lu, C.~N.~Pope and T.~A.~Tran,
  ``Five-dimensional $N{=}4$, SU(2)$\times$U(1) gauged
    supergravity from type IIB,''
  Phys.\ Lett.\  B {\bf 475}, 261 (2000)
  [arXiv:hep-th/9909203].

\bibitem{sound}
G.~Policastro, D.~T.~Son, A.~O.~Starinets,
  ``From AdS/CFT correspondence to hydrodynamics, II. Sound waves,''
  JHEP {\bf 0212}, 054 (2002).
  [hep-th/0210220].

\bibitem{Witten}
  E.~Witten,
  ``Anti-de Sitter space and holography,''
  Adv.\ Theor.\ Math.\ Phys.\  {\bf 2}, 253 (1998)
  [arXiv:hep-th/9802150].

\bibitem{Freedman}
  D.~Z.~Freedman, S.~D.~Mathur, A.~Matusis and L.~Rastelli,
  ``Correlation functions in the CFT($d$)/AdS($d+1$) correspondence,''
  Nucl.\ Phys.\  B {\bf 546}, 96 (1999)
  [arXiv:hep-th/9804058].

\bibitem{Heun}
  R.~S.~Maier,
  ``The 192 Solutions of the Heun Equation'',
  Math.\ Computation {\bf 76}, 811 (2007) [arXiv:math/0408317].

\end{thebibliography}
\end {document}